\documentclass[journal]{new-aiaa} 
\usepackage[utf8]{inputenc}
\usepackage[colorinlistoftodos]{todonotes}
\usepackage{graphicx}
\usepackage{amsmath}
\usepackage[version=4]{mhchem}
\usepackage{siunitx}
\usepackage{caption}
\usepackage{subcaption}
\usepackage{longtable,tabularx}
\usepackage{setspace}
\usepackage{cancel}
\usepackage{float}
\usepackage{multirow}
\usepackage{multicol}
\usepackage{lineno}
\usepackage{draftwatermark}
\SetWatermarkText{For peer review}
\SetWatermarkScale{0.6}
\SetWatermarkLightness{0.9}

\title{Turbulent Accelerating Combusting Flows with a Methane-Vitiated Air Flamelet Model}
\author{Sylvain L. Walsh\footnote{Ph.D student, Department of Mechanical and Aerospace Engineering, AIAA Member; walshsl@uci.edu (Corresponding Author).} and Lei Zhan\footnote{Assistant Specialist, Department of Mechanical and Aerospace Engineering}}
\affil{University of California, Irvine, Irvine, CA, 92697}

\author{Carsten Mehring\footnote{Professor, Leiter Institut für Mechanische Verfahrenstechnik}}
\affil{Universität Stuttgart, Stuttgart, Baden-Württemberg, D-70199, Germany}

\author{Feng Liu\footnote{Professor, Department of Mechanical and Aerospace Engineering, AIAA Fellow.} and William A. Sirignano\footnote{Distinguished Professor, Department of Mechanical and Aerospace Engineering, AIAA Honorary Fellow.}}
\affil{University of California, Irvine, Irvine, CA, 92697}

\begin{document}
\begingroup
\renewcommand{\thefootnote}{}
\footnotetext{Presented as Paper 2024-2427 at the AIAA SciTech 2024 Forum, Orlando, Florida, January 8-12, 2024.}
\endgroup

\maketitle
\begin{abstract}
This work presents a numerical study of a diffusion flame in a reacting, two-dimensional, turbulent, viscous, multi-component, compressible mixing layer subject to a large favorable streamwise pressure gradient. The boundary-layer equations are solved coupled with both the $k$-$\omega$ and SST turbulence models. A compressible extension of the flamelet progress variable method has been proposed and tested for use with large eddy simulations or Reynolds-averaged Navier-Stokes analyses of the burning of methane in pure air and vitiated air; the latter being particularly relevant in turbine burner scenarios. Effects of the level of detail of the reaction mechanism on the sub-grid and resolved-scale computations are studied. A comparison is made with results obtained using a simplified one-step reaction. The numerical results employing the flamelet model with the more detailed reaction mechanism show faster chemistry, significantly reduced peak temperatures and stronger sensitivity to pressure. Vitiated air flames are found to be dominated by unstable solutions, resulting in a weak flame with substantially lower peak temperature and impeded development, struggling to persist without quenching.
\end{abstract}
\section*{Nomenclature}
{\renewcommand\arraystretch{1.0}
\noindent\begin{longtable*}{@{}l @{\quad=\quad} l@{}}
$C$ & progress variable \\
$C_{\mu}$ & $k-\epsilon$ turbulence model empirical constant\\
$C_{\chi}$ &  proportionality constant between turbulence and scalar time-scales\\
$c_p$ & specific heat capacity at constant pressure \\
$h$ & chemical enthalpy\\
$H$ & stagnation enthalpy \\
$k$ & turbulent kinetic energy \\
$N$ & total number of chemical species\\
$p$ & pressure \\
$\mathrm{Pr}$ & Prandtl number\\
$\mathrm{Sc}$ & Schmidt number\\
$T$ & temperature\\
$u$ & velocity in the x-direction \\
$v$ & velocity in the y-direction \\
$x,y$ & cartesian coordinates \\
$Y$ & mass fraction\\
$Z$ & mixture fraction\\
$Z^{''2}$ & variance of the mixture fraction\\
$\epsilon$ & turbulent dissipation rate\\
$\lambda$ & flamelet parameter \\
$\mu$ & mixture-averaged coefficient of viscosity \\
$\rho$ & density \\
$\chi$ & instantaneous scalar dissipation rate\\
$\psi$ & reactive scalar\\
$\omega$ & specific turbulent dissipation rate \\
$\dot{\omega}_C$ & production rate of the progress variable \\
$\widebar{(\cdot)}$ & (overbar) time-averaged or Reynolds-averaged quantity\\
$\widetilde{(\cdot)}$ & (tilde) Favre-averaged value\\
$(\cdot)_{st}$ & stoichiometric composition\\
$(\cdot)_t$ & turbulent quantity\\
\end{longtable*}}

\section{Introduction}
\lettrine{C}{ombustion} in rocket and turbine engines is inherently turbulent due to their accelerating reacting flows with high mass flow rates. Numerically simulating these kinds of processes is difficult because of three main challenges: 1) turbulence
produced by the high-mass-flux reacting flows greatly impacts computational costs due to the wide range of length- and time-scales involved; 2) chemical reactions take place at even smaller length- and time-scales, further increasing computational demands; and 3) the intricate nature of the chemical reactions, involving many chemical species, raises computational demands even further. However, there is significant interest in developing high-fidelity simulations for accelerating reacting flows in the context of new compact designs for gas turbine engines in which the combustion process is extended from the combustion chamber into the turbine passages. 

Thermodynamic analysis performed by \cite{Sirignano1999,Liu2001} shows that this new design, amongst other benefits, allows for: 1) reduced fuel consumption, while increasing specific thrust; 2) widening the operational range of flight Mach number and compressor pressure ratio; and 3) a decrease in pollutant formation and heat transfer losses attributed to the reduction of peak temperatures resulting from flow acceleration. It is important to note that throughout this work, the term "turbine burner" refers to what is known in the literature as the Continuous Turbine Burner (CTB), where combustion occurs within the turbine stages, and the flow is accelerating \cite{Sirignano1999}. This should not be confused with inter-stage turbine burners (ITB), where the CTB cycle performance is approximated by introducing constant-pressure (non-accelerating) burners between turbine stages \cite{Liu2001}. For more information on these design concepts, readers are directed to Yin and Rao’s review \cite{yin_review_2020} on gas turbine engines with turbine burner configurations.

These benefits make augmentative combustion integrated with the turbine function an attractive design feature. Nonetheless, there are major challenges that must be overcome. Specifically, the millisecond residence times and flow acceleration levels of $10^5 g$ pose difficulties for achieving flameholding in the turbine. The latter also increases the likeliness of hydrodynamic instabilities. These instabilities, which include Kelvin-Helmholtz, centrifugal and Rayleigh-Taylor instabilities, have the potential of significantly impacting energy conversion, heat transfer and force loading on the turbine blades. Computational research on reacting mixing layers in accelerating flows is needed for a better understanding of flameholding, mixing efficiency and burning efficiency \cite{Sirignano2012}.

Previous research on compressible, high-speed flows with mixing and chemical reactions is extensive. Miller et al. \cite{miller_experimental_1998} conducted experiments on reacting mixing layers with moderate heat release involving H2-air combustion, aiming to investigate the effects of compressibility on combustion. Day et al. \cite{day1998,day_nonlinear_2001} performed linear and non-linear stability analyses to understand the combined effects of compressibility and heat release on the structure and stability of compressible reacting mixing layers. Ferrer et al. \cite{ferrer_compressibility_2017,martinez_ferrer_compressibility_2017} developed three-dimensional direct numerical simulations (DNS) of high-speed H2-air compressible mixing layers to study growth rates, turbulence characteristics, and stabilization.

To address the specific challenges of burning in the turbine stages of gas turbine engines, where the flow is subjected to strong favorable pressure gradients, our research group has conducted several numerical studies involving accelerating reacting mixing layers. Sirignano et al. \cite{Sirignano1997} developed a theoretical study of a laminar, two-dimensional, compressible, reacting mixing layer that is accelerated due to a pressure gradient in the direction of the flow. They found a similarity solution that reduces the system of partial differential equations (PDEs) to a system of ordinary differential equations (ODEs). Subsequently, Fang et al. \cite{Fang2001} extended this work to encompass more general non-similar cases using a finite-difference method with boundary-layer approximations. Mehring et al. \cite{Mehring2001} advanced these laminar works to include turbulence effects in the boundary-layer equations using a $k-\omega$ turbulence model. Cai et al. \cite{Cai2001,Cai2001_2} studied turbulent transonic accelerating flows by solving the full compressible two-dimensional Navier-Stokes equations. The method was applied to curved ducts, simulating the conditions found in turbines where the flow experiences acceleration due to both streamwise and transverse pressure gradients, thereby providing a more representative analysis of turbine conditions. Cheng et al. \cite{Cheng2007,Cheng2008,Cheng2009} conducted studies on non-premixed combustion in similar accelerating reacting flows. Their focus was on investigating the development of mixing layers from laminar to turbulent flows and transonic regimes. To achieve this, they solved the full two-dimensional Navier-Stokes equations coupled with multi-species equations. Specifically, they examined curved mixing layers passing through turbine-like passages, where it was found that accelerating flows have stabilizing effects while their non-accelerating counterparts do not. Zhu et al. \cite{Zhu2024} conducted a numerical investigation on diffusion flames in transonic flow subjected to large streamwise pressure gradients by solving the steady, compressible, turbulent Navier-Stokes equations using a finite-volume method consisting of a splitting scheme to handle the stiff source terms in the chemical reactions. A notable aspect of their study was the examination of this accelerating reacting flow in a turbine cascade geometry where the inlet stream consisted of vitiated air for a better representation of turbine conditions. They showed that vitiated air has a significant influence on the combustion and aerodynamics. 

Note that all of the aforementioned studies limited their combustion models to single-step/reduced-step reaction mechanisms, which are known to suffer from a loss of detail and accuracy in both the chemistry and the flow. This limitation results in limited accuracy for heat-release predictions and description of ignition and extinction phenomena, which are the governing mechanisms of flame stabilization and flameholding \cite{Pierce2004,Nguyen2018_2}. Therefore, a main goal of the present study is to extend these previous works by employing a combustion model that is more accurate in terms of heat release and ignition/extinction phenomena. Accurately describing these parameters is essential for addressing concerns related to heat transfer to the turbine blades and flameholding in the context of a turbine burner.

The flamelet model, was first introduced by Spalding \cite{SPALDING1971649} and Bilger \cite{bilger1976}, and later further developed by Peters \cite{Peters1984}. It is a widely used method to model sub-scale chemical processes. The model exploits the fact that the sub-grid scales in turbulent combustion describe a regime where local chemical timescales are much faster than the transport timescales, i.e., where values of the second Damköhler number are large. This enables a fast chemistry assumption, wherein the reaction is presumed to occur in asymptotically thin layers called flamelets that are embedded within the turbulent flow. Consequently, turbulent flames can be represented as an ensemble of laminar flame elements (flamelets) that interact with the turbulent flow field. These flamelets, which are usually modeled as counter-flow diffusion flames \cite{Peters1984,Peters2000}, can be solved separately and subsequently coupled to the computational fluid dynamics (CFD) simulation, thereby allowing for the implementation of detailed chemistry including multi-species and multi-step oxidation kinetics, without having to solve for reactive scalars at the resolved scale. Naturally, this method results in considerable computational savings \cite{Nguyen2018}.  

Since the early pioneering work, flamelet models have undergone significant development (see \cite{Cuenot2011}). Notably, Pierce et al. \cite{Pierce2004} developed the flamelet progress variable (FPV) approach to address the oversimplified binary response assumption (burning or non-burning) inherent in classical flamelet models which only consider the stable solutions of the well-known “S-shaped” curve. The FPV approach introduces a reacting scalar, the progress variable $C$, which tracks the global extent of reaction of the mixture. By replacing the instantaneous scalar dissipation rate with the progress variable, the flamelet solutions are parameterized in a way that allows for the unstable branch of the “S-shaped” curve to be contemplated, corresponding to lifted diffusion flames, and lower or non-burning solutions. These lifted diffusion flames are found to occur in turbulent combustion applications and can be expected in turbine burners. Studies such as \cite{Pierce2004,Nguyen2018_2} have shown that better agreement with experiments is achieved when the full “S-shaped” curve is used. Flamelet models based on detailed chemistry and coupled with the FPV approach have been validated against the well-known Purdue experiment. \cite{Nguyen2017,Nguyen2018_2,Nguyen2019,Shadram2021,Shadram2022}.

The original FPV approaches were developed under low-Mach number assumptions. However, these models fail to accurately predict resolved-scale temperatures in the context of compressible high-speed flows, where compressibility effects and viscous heating play an important role. Compressible extensions to FPV approaches have been developed and tested, wherein analytical expressions for resolved-scale temperature are obtained by performing expansions on flamelet-scale mixture properties \cite{pecnik,saghafian_efficient_2015}. In the present study, a formulation for a compressible FPV approach is proposed for use in Reynolds-averaged Navier-Stokes (RANS) and large eddy simulation (LES) analyses that bypasses the need for such expansions. Furthermore, the proposed formulation adds a fourth dimension to the flamelet libraries to account for the dependence of sub-grid chemistry on pressure.

In the present work, the proposed compressible FPV formulation is employed for numerical simulations of accelerating, compressible mixing layers involving the combustion of methane-air and methane-vitiated air, the latter being of particular interest for turbine burners. The obtained results are compared against the studies conducted by Mehring et al. \cite{Mehring2001} and Zhu et al. \cite{Zhu2024} as a means of validation for the proposed formulation.

One of the main benefits of using a flamelet approach is the reduction of computational costs, even when considering intricate reaction mechanisms. However, since the simulations presented here consist of a two-dimensional boundary-layer approximation analysis with minimal computational cost, no attempt to quantify computational savings is made in this paper. These savings become more evident in more complex LES analyses.

This paper is organized as follows: Section \ref{sec:num_frame} presents the computational framework employed. Specifically, Section \ref{gov_eqns} summarizes the governing equations for the transported quantities. Section \ref{sec:comb_model} introduces the proposed compressible FPV formulation and presents sub-scale flamelet results for air and vitiated air. A description of the flow configuration and the numerical solver are given in Sections \ref{sec:flow_conf} and \ref{sec:solver}. In Section \ref{flow_field_results}, resolved-scale computational results are compared to those of Mehring et al. \cite{Mehring2001} and Zhu et al. \cite{Zhu2024}, and the implications of using the proposed FPV formulation are discussed.

\section{Computational Framework}\label{sec:num_frame}
We are concerned with the numerical study of a diffusion flame in a reacting, two-dimensional, turbulent, viscous, multi-component, compressible mixing layer subject to a large favorable streamwise pressure gradient similar to that previously investigated by Mehring et al. \cite{Mehring2001} and Zhu et al. \cite{Zhu2024} using One Step Kinetics (OSK) approaches as combustion models. The flow configuration consists of two streams, namely a pure or vitiated air stream and a fuel vapor stream, initially separated by a flat plate with slip walls. The computational domain extends downstream from the trailing edge of the flat plate, where the two streams come into contact and chemical reactions occur between the reactants, leading to the formation of a diffusion flame near the location of stoichiometric composition. 
\subsection{Governing Equations}\label{gov_eqns}
Applying the boundary-layer approximation to the Favre-averaged, steady, compressible Navier-Stokes equations for a multi-component reacting flow gives the following equations for axial momentum and stagnation enthalpy \cite{Mehring2001}:

\begin{equation}\label{eq:mom}
\frac{\partial \tilde{u}}{\partial x}=-\frac{\tilde{v}}{\tilde{u}}\frac{\partial\tilde{u}}{\partial y}-\frac{1}{\bar{\rho}\tilde{u}}\frac{d\bar{p}}{dx}+\frac{1}{\bar{\rho}\tilde{u}}\frac{\partial}{\partial y}\left[(\mu+\mu_t)\frac{\partial\tilde{u}}{\partial y}\right]
\end{equation}
\begin{equation}\label{eq:energy}
\begin{split}
    \frac{\partial H}{\partial x}=-\frac{\tilde{v}}{\tilde{u}}\frac{\partial H}{\partial y}+\frac{1}{\bar{\rho}\tilde{u}}\frac{\partial}{\partial y}\left[\left(\frac{\mu}{\mathrm{Pr}}+\frac{\mu_t}{\mathrm{Pr_t}}\right)\frac{\partial H}{\partial y}\right]+\frac{1}{\bar{\rho}\tilde{u}}\frac{\partial}{\partial y}\left[(\mu+\sigma^*\mu_t)\frac{\partial k}{\partial y}\right]\\
    +\frac{1}{\bar{\rho}\tilde{u}}\frac{\partial}{\partial y}\left[\left\{\mu\left(\frac{1}{\mathrm{Sc}}-\frac{1}{\mathrm{Pr}}\right)+\mu_t\left(\frac{1}{\mathrm{Sc_t}}-\frac{1}{\mathrm{Pr_t}}\right)\right\}\sum_{i=1}^N\tilde{h}_{i,t}\frac{\partial \widetilde{Y}_i}{\partial y}\right]\\
    +\frac{1}{\bar{\rho}\tilde{u}}\frac{\partial}{\partial y}\left[\left\{\mu\left(1-\frac{1}{\mathrm{Pr}}\right)+\mu_t\left(1-\frac{1}{\mathrm{Pr_t}}\right)\right\}\frac{\partial(\tilde{u}^2/2)}{\partial y}\right]
\end{split}  
\end{equation}
where the stagnation enthalpy is defined as:
\begin{equation}\label{eq:energy_components}
    H=\tilde{h}+\frac{1}{2}\tilde{u}^2+k
\end{equation}
with $k$ being the turbulent kinetic energy (t.k.e.). $\sigma^*$ in Eq. \ref{eq:energy} is a constant pertaining to the two-equation turbulence models which will be covered later. The average total chemical enthalpy and total chemical enthalpy of species $i$ are given by
\begin{equation}
    \tilde{h}=\sum_{i=1}^{N}\widetilde{Y}_i\tilde{h}_{i}
\end{equation}
and
\begin{equation}\label{eq:chemical_energy}
    \tilde{h}_{i}=\int_{T_{\mathrm{ref}}}^{\widetilde{T}}c_{p,i}dT'+\Delta h_i^o(T_{\mathrm{ref}})
\end{equation}
respectively, where $c_{p,i}$ and $\Delta h_i^o$ are the specific heat at constant pressure and the heat of formation of species $i$ at temperature $T_{\mathrm{ref}}$, and $N$ denotes the number of species considered in the reaction mechanism. The specific heats are functions of temperature and determined by NASA empirical polynomials \cite{Mcbride1993}.

Mixture-averaged dynamic viscosity $\mu$, thermal conductivity and molecular diffusivity are computed with the Lennard-Jones inter-molecular potential model, and verified with the Cantera chemical kinetics software \cite{cantera}. Under the restrictions of Fickian diffusion, molecular diffusivity is computed using self-diffusion coefficients. Consequently, Schmidt and Prandtl numbers can be determined using mixture-averaged molecular properties, allowing for departure from unity Lewis number assumptions.
$\mathrm{Sc_t}$ and $\mathrm{Pr_t}$ are the turbulent Schmidt and Prandtl numbers. In the present study, their values are assumed to be constant and are set to $\mathrm{Sc_t}=\mathrm{Pr_t}=1.0$. 
Furthermore, the continuity equation is used to determine the transverse component of velocity.

Note that transport equations for species conservation are not given, as the species mass fractions will appear through the flamelet equations. Species mass fractions, which are stored in library form, are provided by the flamelet model. Equations (\ref{trans_z}) through (\ref{trans_c}) below are transport equations (Ref. \cite{pecnik}) for the Favre mean mixture fraction $\widetilde{Z}$, variance of the mixture fraction $\widetilde{Z''^2}$ and progress variable $\widetilde{C}$, which all serve as lookup parameters for the flamelet libraries. That is, given a combination of $\widetilde{Z}$, $\widetilde{Z''^2}$, and $\widetilde{C}$ at each computational cell, species mass fractions  which were pre-computed by solving the system of flamelet equations may be determined by interpolating the flamelet library. 
\begin{equation}\label{trans_z}
    \frac{\partial\widetilde{Z}}{\partial x}=-\frac{\tilde{v}}{\tilde{u}}\frac{\partial{\widetilde{Z}}}{\partial y}+\frac{1}{\bar{\rho}\tilde{u}}\frac{\partial}{\partial y}\left[\left(\frac{\mu}{\mathrm{Sc}}+\frac{\mu_t}{\mathrm{Sc_t}}\right)\frac{\partial\widetilde{Z}}{\partial y}\right]
\end{equation}
\begin{equation}\label{trans_varz}
   \frac{\partial\widetilde{Z''^2}}{\partial x}=-\frac{\tilde{v}}{\tilde{u}}\frac{\partial{\widetilde{Z''^2}}}{\partial y}+\frac{1}{\bar{\rho}\tilde{u}}\frac{\partial}{\partial y}\left[\left(\frac{\mu}{\mathrm{Sc}}+\frac{\mu_t}{\mathrm{Sc_t}}\right)\frac{\partial\widetilde{Z''^2}}{\partial y}\right]+\frac{2}{\bar{\rho}\tilde{u}}\frac{\mu_t}{\mathrm{Sc_t}}\left(\frac{\partial\widetilde{Z}}{\partial y}\right)^2-\frac{1}{\tilde{u}}\widetilde{\chi}
\end{equation}
\begin{equation}\label{trans_c}
\frac{\partial\widetilde{C}}{\partial x}=-\frac{\tilde{v}}{\tilde{u}}\frac{\partial{\widetilde{C}}}{\partial y}+\frac{1}{\bar{\rho}\tilde{u}}\frac{\partial}{\partial y}\left[\left(\frac{\mu}{\mathrm{Sc}}+\frac{\mu_t}{\mathrm{Sc_t}}\right)\frac{\partial\widetilde{C}}{\partial y}\right]+\frac{1}{\bar{\rho}\tilde{u}}\widetilde{\dot{\omega}}_C
\end{equation}
The last term on the r.h.s of Eq. \ref{trans_varz} is the mean scalar dissipation rate and is frequently modeled as $\widetilde{\chi}=C_{\chi}\frac{\tilde{\epsilon}}{\tilde{k}}\widetilde{Z''^2}=C_{\chi}C_{\mu}\widetilde{\omega}\widetilde{Z''^2}$ 
\cite{Peters1984,Peters2000} with $k$, $\epsilon$ and $\omega$ being the turbulent kinetic energy, mean turbulent dissipation rate and specific turbulent dissipation rate, respectively \cite{Peters2000}. $C_{\chi}$ is the proportionality constant between turbulence and scalar time-scales. In the present study, it is set to $C_{\chi}=2.0$, as this value was found to adequately predict the decay of scalar variance in an inert jet of methane \cite{Janicka1982}. $C_{\mu}$ is an empirical constant set to 0.09 and relates $k$ and $\epsilon$ to the turbulent viscosity in the Chien $k-\epsilon$ turbulence model \cite{Chien1982}. The source term $\widetilde{\dot{\omega}}_C$ in Eq. \ref{trans_c} is the production rate of the progress variable which is obtained from the flamelet library. It is defined as the summation of the reaction rates for the individual species involved in the definition of the progress variable within the flamelet libraries. 

Once species mass fractions are available, temperature is determined from the resolved-scale energy equation. Specifically, temperature is derived from incremental changes in sensible enthalpy determined through Eqs. (\ref{eq:energy_components}) and (\ref{eq:chemical_energy}), with $c_p$ being mixture-averaged by the mass fractions provided by the flamelet libraries, and evaluated using the resolved-scale mean temperature. Then, since pressure is prescribed, the ideal gas equation of state yields density.

Regarding the turbulent viscosity, Mehring et al. \cite{Mehring2001} employed Wilcox's 1988 two-equation $k-\omega$ model, incorporating varying closure coefficients to address compressibility effects \cite{Wilcox1988,wilcox_book}. Recognizing the sensitivity of this model to freestream turbulence levels \cite{Menter1992} and anticipating the use of the current flamelet model in Detached Eddy Simulations (DES) in future studies, the current work employs both the $k-\omega$ model and Menter et al.'s 2003 improved $k-\omega$ Shear-Stress Transport (SST) model \cite{menter2003}. Given the absence of a wall in the current computations, the values of the SST blending function ($F_1$ and $F_2$ within the model) are set to zero. Note that the closure coefficients employed here do not deviate from those specified in the references.

\subsection{Turbulent Combustion Model}\label{sec:comb_model}
The turbulent combustion model employed is an extension of the low Mach number flamelet progress variable approach (FPV) \cite{Peters1984,Peters2000,Pierce2004} for use in compressible high-speed flows such as the ones being studied here. In this approach, flamelet libraries are generated under specific reactant composition and temperature boundary conditions that match the resolved-scale boundary conditions of Mehring et al. \cite{Mehring2001} and Zhu et al. \cite{Zhu2024}. The fuel stream is composed of pure $\mathrm{CH_4}$ vapor at 400 K. As for the oxidizer stream, both pure air and vitiated air compositions are considered. The latter intends to better approximate the conditions within a turbine burner: the inlet stream of air entering the turbine passages is composed of a mixture of air and burned gases that result from the fuel-lean combustion occurring in the combustion chamber upstream of the turbine. It is estimated that a fuel-to-air ratio of 0.03 in the primary combustion chamber is required in order to produce a turbine inlet temperature of 1650 K, which results in a vitiated air oxidizer stream composition of 73.77\% $\mathrm{N_2}$, 11.01\% $\mathrm{O_2}$, 8.04\% $\mathrm{CO_2}$ and 7.18\% $\mathrm{H_2O}$ by mass. The oxidizer temperature and composition for the pure air library consists of 1650 K and 76\% $\mathrm{N_2}$, 23.37\% $\mathrm{O_2}$, and 0.63\% $\mathrm{H_2O}$ by mass.

The chemical reaction mechanism utilized for the flamelet library generation is Version 1.0 of the Foundational Fuel Chemistry Model (FFCM-1) \cite{Smith2016,TAO201818}. The FFCM-1 mechanism, which consists of 38 species and 291 elementary reactions, incorporates up-to-date kinetic knowledge with well-defined predictive uncertainties for small hydrocarbon fuels. The mechanism has been validated against experimental data for methane-oxygen combustion over a pressure range of 10-100 atm, making it well suited for the current study. Owing to the objective of studying the impact of the detail of chemical mechanism on the resolved-scale computations, flamelet libraries are also generated using a 13-species, 32-elementary reaction, skeletal reduction of the FFCM-1 mechanism. The species considered in the skeletal reduction are as follows: $\mathrm{H_2}$, $\mathrm{H}$, $\mathrm{O_2}$, $\mathrm{O}$, $\mathrm{OH}$, $\mathrm{HO_2}$, $\mathrm{H_2O}$, $\mathrm{CH_3}$, $\mathrm{CH_4}$, $\mathrm{CO}$, $\mathrm{CO_2}$, $\mathrm{CH_2O}$ and $\mathrm{N_2}$. Nitrogen is treated as inert, meaning it participates in elementary reactions only as a third body. This 13-species skeletal mechanism has been used by Zhan et al. \cite{zhan_combustion_2024} in a FPV approach for DES simulations to study combustion dynamics of a ten-injector liquid rocket engine.  

\subsubsection{Determination of resolved-scale temperature}
The FlameMaster code \cite{flamemaster} is employed to solve the system of steady flamelet equations in a counter-flow configuration \cite{Peters1984}, namely,
\begin{equation}\label{eq:steady_flamelet}
    -\rho \frac{\chi(Z)}{2}\frac{\partial^2 \psi_j}{\partial Z^2}=\dot{\omega}_j, \hspace{1mm}j=1,2,...,N+1
\end{equation}
    where $Z$ is the mixture fraction and $\psi_j$ represents the reactive scalars, comprised of the $N$ species mass fractions and temperature. The chemical source terms $\dot{\omega}_j$, correspond to reaction rates for species mass fraction equations and heat-release rate for the temperature equation. The instantaneous scalar dissipation rate, denoted as $\chi$, describes the local effect of molecular mixing on the chemistry and implicitly incorporates the interaction between turbulence and combustion. It has units of 1/sec and is directly proportional to the flamelet ambient strain rate. A functional form of $\chi(Z)$ is derived from the application of flamelet equations in a counter-flow potential flow configuration \cite{Peters2000}. This geometry, while different from the resolved-scale configuration, is used solely for the sub-scale computations, specifically for solving the flamelet equations.

    Note that the flamelet temperature equation in Eq. \ref{eq:steady_flamelet} is obtained by neglecting the viscous dissipation term and the exchange between kinetic energy and internal energy due to compressibility, based on the low Mach number approximation. Additionally, the $\partial p/\partial t$ term and radiative heat transfer are omitted \cite{Peters1984}. These assumptions do not hold for the high-speed expanding flow within the turbine, where compressibility effects and viscous heating play an important role. To address this issue, compressible variants of the flamelet progress variable approach have been developed and tested, where temperature is computed from the resolved-scale energy equation and the tabulated species mass fractions \cite{pecnik,saghafian_efficient_2015}. In these approaches, analytical expressions for resolved-scale temperature are obtained by performing expansions on flamelet-scale mixture properties, such as heat capacity, to account for deviations due to the differences between resolved-scale and flamelet temperatures. These methods heavily rely on the assumption that the resolved-scale species mass fractions do not vary significantly from the sub-scale flamelet species mass fractions. This assumption is valid on the condition that the deviation in temperature obtained between the resolved-scale energy equation and the flamelet temperature equation is not too large. Saghafian et al. \cite{saghafian_efficient_2015} showed that the assumption holds well for major species when applying temperature deviations of up to 400 K and 1000 K in fuel and oxidizer temperature, respectively. Minor species were found to deviate more from their nominal values; however, their contributions to the mixture properties are much smaller than those of the major species.

    Making use of the aforementioned assumption that the resolved-scale species mass fractions can be considered equal to the sub-grid species mass fractions, in this work, we propose an alternative method whereby temperature is obtained directly from the resolved-energy equation using the interpolated species mass fractions from the library, bypassing the need for the analytical temperature equation. Here, instead of expanding mixture properties from the libraries, $\widetilde{c_p}$ is determined using the resolved-scale temperature, as detailed in Section \ref{gov_eqns}. The benefit of this approach is two-fold: 1) a reduction in the size of the flamelet libraries, as mixture properties are computed on the resolved scale and do not need to be stored in the libraries; and 2) the mixture-averaged heat capacity, and consequently temperature, are determined directly in accordance with the resolved-scale energy equation.
    
\begin{figure}
\centering
\begin{subfigure}{.5\textwidth}
  \centering
  \includegraphics[]{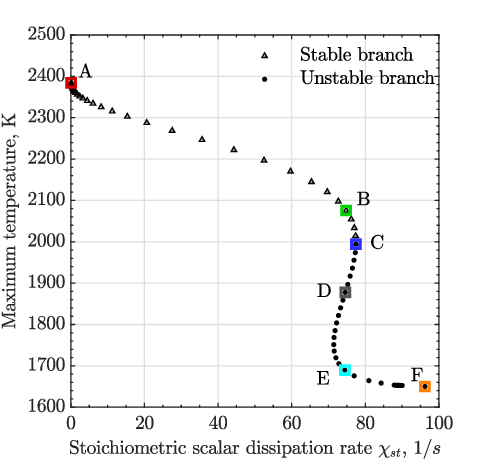}
  \caption{Maximum temperature versus $\chi_{st}$.}
  \label{fig:s_vit_air}
\end{subfigure}%
\begin{subfigure}{.5\textwidth}
  \centering
  \includegraphics[]{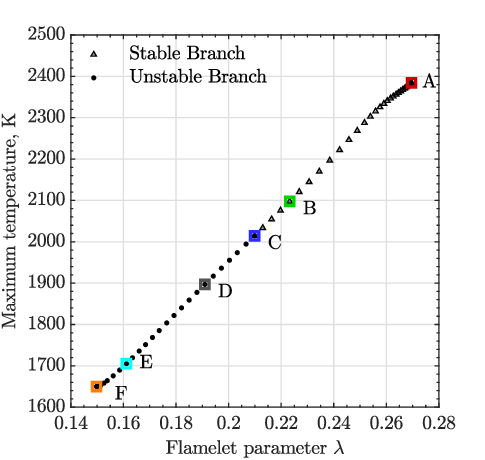}
  \caption{Maximum temperature versus $\lambda$.}
  \label{fig:spv_vit_air}
\end{subfigure}
\caption{Vitiated air steady flamelet solution space using the FFCM-13 mechanism and a background pressure of 30 bar.}
\label{fig:s-and-pv}
\end{figure}

\subsubsection{Flamelet progress variable approach}
The computed laminar flamelet solutions using vitiated air and the FFCM-13 mechanism at a background pressure of 30 bar are presented in Fig. \ref{fig:s_vit_air} as the "S-shaped" curve, where the maximum temperature is plotted against $\chi_{st}$, the scalar dissipation rate at stoichiometric $Z$ value. Each point represents a flamelet solution consisting of reactive scalar profiles parameterized by the mixture fraction. Starting from the upper left corner of Fig. \ref{fig:s_vit_air} ($\chi_{st}\rightarrow 0$), the fully stable equilibrium solution (point A) is observed. As the strain rate (scalar dissipation rate) increases, the flame temperature decreases until reaching the quenching limit at point C. This temperature drop results from increased heat conduction from the reaction zone to the outer nonreacting zones. Typically, further increasing strain beyond point C leads to the critical point where heat conduction from the reaction zone outweighs heat production by the chemical reaction, causing the flamelet to quench. However, due to the high oxidizer temperature (1650 K), "marginally burning" solutions are encountered beyond $\chi_{st}=80$ 1/s. If both the strain rate and temperature are decreased from the quenching limit point C, unstable solutions are encountered. The right-most point along the unstable branch (point F) corresponds to the non-burning flamelet solution, where reactants simply diffuse into each other without chemical reaction.

This behavior results in multi-valued solutions, meaning that for a given value of $\chi_{st}$ (e.g., points B, D, and E), multiple solutions with varying maximum temperature and species compositions are encountered. These include a solution on the stable branch (point B), a solution on the unstable branch (point D), and a solution on the "marginally burning" stable branch (point E). It has been shown \cite{Pierce2004,Nguyen2018_2} that including both the unstable and stable burning branches more accurately reflects the extinction, re-ignition, and associated flameholding that occurs in dynamic situations. This motivates the use of the FPV for the current work, whereby flamelet solutions are monotonically mapped to a flamelet parameter $\lambda$ defined as
\begin{equation}
    \lambda = C(Z_0)
\end{equation}
where $C$ is the progress variable and $Z_0$ is a reference value of Z. The progress variable is typically defined as a linear combination of the mass fractions of a chosen set of representative products. The choice of $Z_0$ and the products that comprise $C$ is ad hoc and will differ depending on the reactants and boundary conditions for temperature and reactant composition. However, the selection must guarantee the bijective mapping of the reactive scalars to $\lambda$, such they can be uniquely described by 
\begin{equation}\label{eq:steady_soln}
    \psi_j=\psi_j(Z,\lambda)
\end{equation}
In the present work, $Z_0$ is chosen to be the stoichiometric mixture fraction $Z_{st}$, and $\mathrm{H_2O}$ and $\mathrm{CO_2}$ are used for the definition of $C$. Note that the source term of the progress variable (right-most term in Eq. \ref{trans_c}) is also defined in this step with the same species and weights. Figure \ref{fig:spv_vit_air} shows the bijective mapping of the vitiated air flamelets to $\lambda$, where maximum temperature monotonically increases from points F to A. Similar behavior is observed with the solutions generated using the full FFCM-1 mechanism and for solutions involving pure air as the oxidizer.

\subsubsection{Pressure dimension of the flamelet libraries}
\begin{figure}
    \centering
    \includegraphics[]{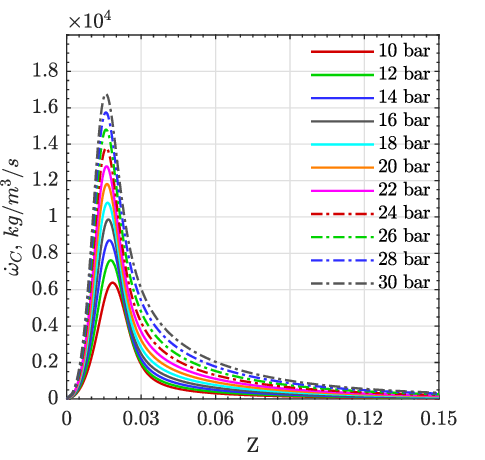}
    \caption{${\dot{\omega}_C}$ vs Z.}
    \label{fig:omc_vs_p}
\end{figure}

In the context of high-speed compressible flows, the production rate of the progress variable $\dot{\omega}_C$ is highly sensitive to compressibility effects, particularly due to the impact of pressure on unimolecular elementary reactions. Given the rapidly expanding flows studied here, it is crucial to consider the dependence of $\dot{\omega}_C$ on background pressure. Figure \ref{fig:omc_vs_p} illustrates quenching-limit (point C in Fig. \ref{fig:s-and-pv}) $\dot{\omega}_C$ profiles as a function of the mixture fraction $Z$ for various background pressures, clearly demonstrating the strong pressure dependence. The production rate $\dot{\omega}_C$ peaks near the stoichiometric mixture fraction ($Z{st}=0.027$ for this vitiated air composition) and increases monotonically with rising background pressure.

Previous studies have employed power-laws to rescale the source term of the progress variable to account for their dependence on background pressure \cite{pecnik,saghafian_efficient_2015,zhan_combustion_2024}. In the current work, we bypass the use of power-laws by adding a dimension to the flamelet libraries to account for variations in background pressure. Specifically, reactive scalars are determined from Eq. \ref{eq:steady_flamelet} for background pressures ranging from 10 to 30 bar and tabulated as 
\begin{equation}\label{eq:steady_soln_p}
    \psi_j=\psi_j(Z,\lambda,p)
\end{equation}

\begin{figure}
\centering
\begin{subfigure}{.5\textwidth}
  \centering
  \includegraphics[]{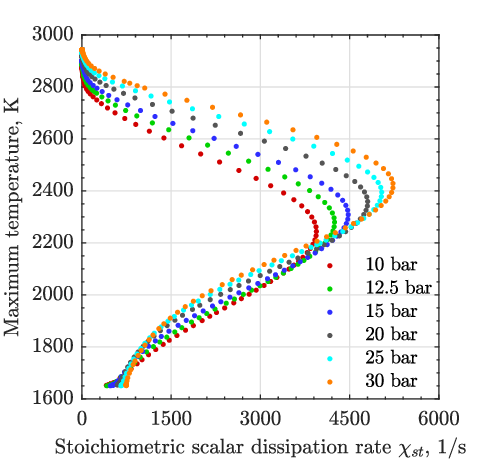}
  \caption{Pure air.}
  \label{fig:s_shaped_pure_air}
\end{subfigure}%
\begin{subfigure}{.5\textwidth}
  \centering
  \includegraphics[]{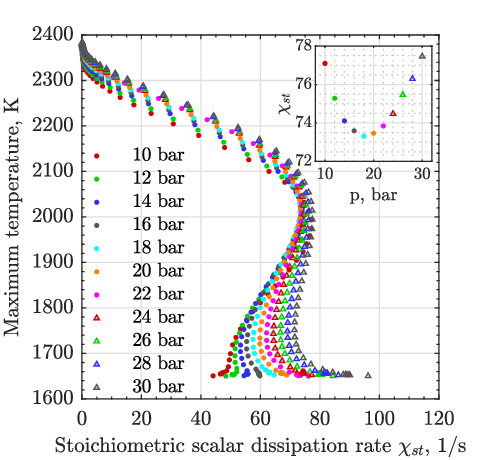}
  \caption{Vitiated air.}
  \label{fig:s_shaped_vitiated_air}
\end{subfigure}
\caption{Steady flamelet solutions represented as "S-shaped" curves for different values of background pressure.}
\label{fig:s_shaped_new}
\end{figure}

Figure \ref{fig:s_shaped_new} presents "S-shaped" curves for various background pressures. In the case of pure air (Fig. \ref{fig:s_shaped_pure_air}), both the peak temperature and quenching limits increase monotonically with rising background pressure for most of the solution space, with quenching limits increasing by up to 25\% with respect ti the lowest background pressure value. The stable branch of these solutions show a stronger dependence on background pressure. Conversely, in the vitiated air case (Fig. \ref{fig:s_shaped_vitiated_air}), the unstable solutions exhibit a stronger dependence on background pressure. More pressure evaluation points have been used in this case due to the prevalence of unstable solutions observed in the resolved-scale computations (as detailed in Section \ref{flow_field_results}). Notably, for the vitiated air composition, the quenching limit does not increase monotonically with background pressure but instead follows a quadratic trend, as seen in the top-right sub-figure. This phenomenon cannot be captured using power-law methods, highlighting the advantage of the current approach for simulations involving vitiated air. Furthermore, the vitiated air solutions show quenching limits two orders of magnitude smaller than their pure air counterparts. This is anticipated to have a notable effect on the development of the vitiated air flames in resolved-scale computations.

Given the influence of pressure on flammability limits and on the characteristics of unstable branches, especially in the case of vitiated air, these findings further underscore the necessity for a combustion model capable of accurately describing these dependencies on pressure. Such precision is crucial for understanding flameholding and the dynamics of ignition and extinction.

\subsubsection{Coupling of the flamelet libraries to the resolved-scale parameters}
In order to couple the quasi-laminar solutions to the resolved-scale computations, Favre-averaged means of the reactive scalars are determined as first moments through the presumed shape PDF approach,
\begin{equation}\label{eq:lambda_mean}
\widetilde{\psi}_j=\widetilde{\psi}_j(\widetilde{Z},\widetilde{Z''^2},\lambda,\bar{p})=\int_0^1 \psi_j(Z,\lambda,\bar{p})\widetilde{P}(Z,\widetilde{Z},\widetilde{Z''^2})dZ
\end{equation}
Equation \ref{eq:lambda_mean} is obtained by involving the following standard assumptions: 1) the mixture fraction is assumed to be independent of pressure \cite{pecnik,urzay2012stochastic}; 2) the marginal probability of pressure is a delta function that centers the integral on the resolved-scale average pressure \cite{urzay2012stochastic}; 3) The marginal distribution of the mixture fraction is given by a beta-PDF which is parameterized by the mean mixture fraction $\widetilde{Z}$ and the mean variance of the mixture fraction $\widetilde{Z''^2}$ \cite{cook_subgrid_1994,jimenez_priori_1997,wall_evaluation_2000}; 4) if the flamelet parameter $\lambda$ is defined to be independent of the mixture fraction and pressure, the conditional PDF for $\lambda$, given mixture fraction and pressure, can also be described by a delta function \cite{janicka_two-variables_1979,Pierce2004}. For more details on the presumed shape PDF approach employed in the current work, readers are directed to \cite{walsh_turbulent_2024}.

To acquire mean reactive scalars from the flamelet library, all that remains is to find a functional relationship between the flamelet parameter $\lambda$ and the resolved scale mean progress variable $\widetilde{C}$, such that, given a combination of $\widetilde{Z}$, $\widetilde{Z''^2}$, and $\widetilde{C}$ at each computational cell, reactive scalars may be determined by interpolating the flamelet library 
\begin{equation}
\widetilde{\psi}_j=\widetilde{\psi}_j(\widetilde{Z},\widetilde{Z''^2},\widetilde{C},\bar{p})
\end{equation}
 Pierce et al. \cite{Pierce2004} proposed that this functional relationship should be established by computing $\widetilde{C}$ as the first moment of the laminar progress variable profiles, $C=C(Z,\lambda,\bar{p})$, using Eq. \ref{eq:lambda_mean}. In this context, if an appropriate definition has been given to $C$, $\lambda$ is uniquely related to $\widetilde{C}$ for each combination of $\widetilde{Z}$ and $\widetilde{Z''^2}$.
 
In order to streamline the generation of turbulent flamelet libraries, this study simplifies the relationship between $\lambda$ and $\widetilde{C}$ by setting them equal to each other, i.e., $\lambda = \widetilde{C}$. It is acknowledged that such a simplification introduces inconsistencies that may lead to inaccuracies, including: 1) neglecting turbulent fluctuations in the progress variable by equating the mean progress variable to its laminar counterpart $\lambda$; and 2) the resolved-scale transport equation for the progress variable tracks a linear combination of product mass fractions for any value of the mixture fraction, rather than specifically the stoichiometric mixture fraction (recall the definition of $\lambda$). While this approach has been used before \cite{Nguyen2017,Nguyen2018,Nguyen2018_2,Nguyen2019,Shadram2021,Shadram2022,zhan_combustion_2024}, in this work, we seek to quantify the accuracy of this method. It is demonstrated that, in and near the reaction zone, the simplification $\lambda = \widetilde{C}$ yields reactive scalars that differ only marginally from those obtained when considering the functional relationship between $\lambda$ and $\widetilde{C}$ as determined by the first moment of the progress variable.

\begin{figure}
    \centering
    \includegraphics[]{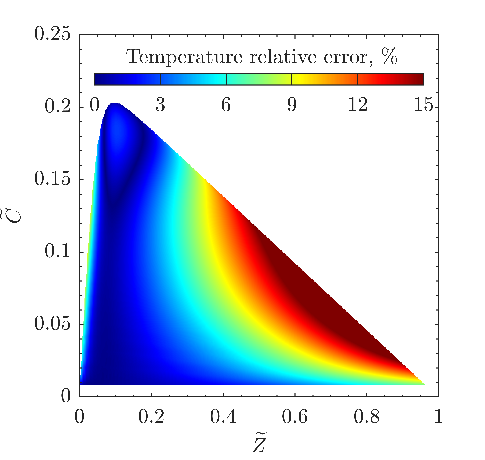}
    \caption{Relative error in mean flame temperature when using the simplification $\lambda=\widetilde{C}$.}
    \label{fig:consistency_error}
\end{figure}

Figure \ref{fig:consistency_error} shows contours representing the relative error in mean flame temperature resulting from the simplification $\lambda=\widetilde{C}$ compared to the more intricate functional relationship proposed by \cite{Pierce2004}. This error is assessed with the pure air libraries generated using the FFCM-13 mechanism at a constant background pressure of 30 bar and a fixed value for $\widetilde{Z''^2}$ of 0.1. Throughout the reaction zone, represented by the dark blue area surrounding $Z_{st}=0.055$ (for pure air), the relative temperature error remains below 5\%, ensuring adequate accuracy in the flame region for resolved-scale computations. Similar trends are observed for species mass fractions and the production rate of the progress variable $\widetilde{\dot{\omega}}_C$ across different pressures and expected values of $\widetilde{Z''^2}$, as well as with both the skeletal and full reaction mechanisms.

\subsubsection{Comparison of FFCM-13 and FFCM-1 reaction mechanisms}\label{sec:ffcm1_ffcm13_flamelet}

As previously discussed, flamelet libraries have been generated for both the FFCM-13 (skeletal/reduced) and FFCM-1 (full) reaction mechanisms. When reducing a mechanism, it is inevitable that some error will be introduced due to the alteration of reaction pathways when certain species and reactions are omitted. However, mechanism reduction within a flamelet progress variable approach offers two significant advantages: 1) reduced storage requirements for the flamelet libraries, and 2) more importantly, decreased access time for reactive scalars during CFD simulations. It is important to note that for every resolved-scale computational cell, the 4D flamelet libraries must be interpolated to retrieve each reactive scalar. Consequently, the library access time is proportional to the number of chemical species considered. Utilizing the reduced FFCM-13 mechanism results in approximately a two-thirds reduction in both storage requirements and access time. Thus, there is justification for using the simplest mechanism possible without compromising key physical phenomena.

\begin{figure}
    \centering
    \includegraphics[]{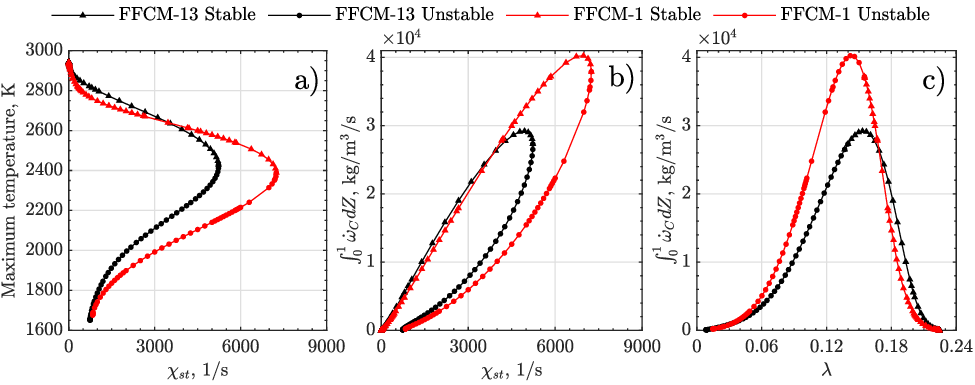}
    \caption{Comparison between sub-scale results obtained with both the FFCM-13 (skeletal) and FFCM-1 (full) reaction mechanisms at a background pressure of 30 bar.}
    \label{fig:ffcm_vs_ffcm13}
\end{figure}

To assess the resulting differences, sub-scale results from both mechanisms are compared in Fig. \ref{fig:ffcm_vs_ffcm13},
at a background pressure of 30 bar with pure air on the oxidizer side of the counter-flow configuration (flamelet). Variations between the two sets are anticipated due to differences in reaction pathways, with one mechanism involving 13 species and 38 reactions, and the other involving 38 species and 291 reactions. 

Figure \ref{fig:ffcm_vs_ffcm13}a) shows the maximum flamelet temperature against the instantaneous stoichiometric scalar dissipation rate, the latter being proportional to the ambient strain rate. The counter-flow diffusion flame using the FFCM-1 reaction mechanism can endure higher strain rates before extinguishing, with a roughly 40\% increase in the maximum $\chi_{st}$ compared to the FFCM-13 results. At higher strain rates, the residence time is reduced, and the higher quenching limit of the FFCM-1 mechanism indicates faster chemistry compared to its skeletal counterpart. Similar results are observed for vitiated air, where the increase in the quenching limit is even greater, approximately double that of the skeletal mechanism. Additionally, for most of the shared solution space (i.e., points with the same $\chi_{st}$), the FFCM-13 mechanism yields higher peak temperatures. This behavior is expected, as the reduced mechanism generates more heat release due to less dissociation compared to the full mechanism. As a result, higher temperatures are anticipated in resolved-scale computations when using the skeletal mechanism. Nevertheless, solutions along the stable branch show better agreement, particularly those near the equilibrium solution (upper left-most points).

The integrated production rate of the progress variable, $\dot{\omega}_C$, plotted against $\chi_{st}$ for both mechanisms is shown in Fig. \ref{fig:ffcm_vs_ffcm13}b). Here, the integration is performed over $Z$. Recall that $\dot{\omega}_C$ is defined as the sum of the reaction rates of $\mathrm{CO_2}$ and $\mathrm{H_2O}$. For both mechanisms, the production rate peaks along the stable branches near the quenching limits, scaling mostly monotonically with ambient strain rate. This behavior is expected, as a lower ambient strain rate reduces diffusion, which drives the reaction. Furthermore, a reduction in strain rate increases the flamelet residence time, leading to a higher volume. Given the higher maximum tolerable strain rate of the FFCM-1 mechanism, it is expected to exhibit larger values of integrated progress variable production. Figure \ref{fig:ffcm_vs_ffcm13}c) shows the same integrated production term, but plotted against the flamelet parameter $\lambda$, which is related to the resolved-scale mean progress variable as $\lambda=\widetilde{C}$. Aside from the higher maximum value for the detailed mechanism, it is observed that along the entirety of the unstable branches and parts of the stable branches, the FFCM-1 mechanism has higher production rates than the skeletal mechanism for the same values of $\lambda$. Since the growth of $\widetilde{C}$ at the resolved-scale (which governs the resolved-scale reaction) is driven by the magnitude of $\dot{\omega}_C$ (see Eqn. \ref{trans_c}), we can conclude that the FFCM-1 (full) mechanism should ignite sooner.

Two important findings emerge from these comparisons: 1) the skeletal mechanism is likely to over-predict resolved-scale temperatures, and 2) the significant differences in quenching strain rates suggest that ignition may be notably affected when using the reduced mechanism. However, whether these differences will influence the resolved-scale computations remains to be verified, which will be addressed in Section \ref{rs_mech_comp}.

\subsection{Flow Configuration}\label{sec:flow_conf}

\begin{figure}
    \centering
    \includegraphics[]{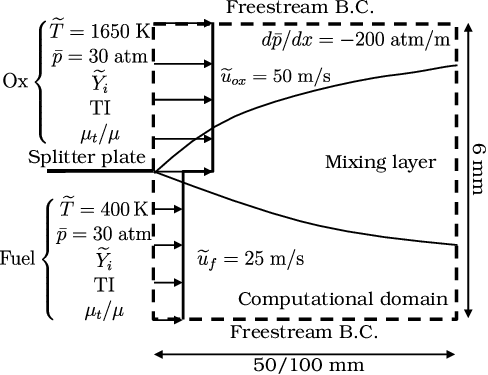}
    \caption{Diagram of the flow configuration.}
    \label{fig:flow_config}
\end{figure}

The flow configuration in this study, illustrated in Fig. \ref{fig:flow_config}, which replicates the boundary-layer analysis of Mehring et al. \cite{Mehring2001} and the two-dimensional Navier-Stokes analysis of Zhu et al. \cite{Zhu2024}, serves as a simplified model for combustion in turbine passages. The computational domain is rectangular, with a width of either 50 mm or 100 mm, depending on the case, and a height of 6 mm. The left boundary, located at the trailing edge of the splitter plate (at x = 0.0 mm), serves as the inlet for two separate reactant streams. Since, the splitter plate is outside of the computational domain, wall effects are not considered. 

The upper stream is a hot air mixture at 1650 K with a uniform streamwise velocity of 50 m/s. This hot air is either pure or vitiated, with its compositions detailed in Section \ref{sec:comb_model}. The lower stream consists of pure $\mathrm{CH_4}$ vapor at 400 K, with a uniform streamwise velocity of 25 m/s. Freestream boundary conditions are applied at both the top and bottom edges of the domain. The inlet pressure for both streams is $\bar{p}$ = 30 atm. To simulate the rapidly expanding flows within a turbine, a strong favorable pressure gradient of 200 atm/m in the positive downstream (x) direction is applied, consistent with previous studies \cite{Mehring2001,Zhu2024}. This specific value is a “ballpark” estimate of the average pressure gradient observed by the flow within a turbine section and is reasonable for a gas turbine engine operating at sea level with a turbine inlet pressure of 30 atm and a turbine passage length of roughly 10-20 cm. 

As described in Section \ref{gov_eqns}, both the $k-\omega$ and SST two equation turbulence models are used. For simulations using the $k-\omega$ model, the inlet turbulent intensity (TI) and the ratio of turbulent to molecular viscosity ($\mu_t/\mu$) are set to 2.58E-2\% and 5.40E-2, respectively, for the oxidizer stream, and 5.16E-2\% and 5.10E-1 for the fuel stream. These values match the freestream $k$ and $\omega$ values used by Mehring et al. \cite{Mehring2001}, which were selected for validation of non-reacting computations against experimental data (see Fig. 1 in \cite{Mehring2001}). Since turbine inlet flows are expected to have higher turbulence levels, the turbulence intensity is increased for simulations using the SST model. In these cases, TI and $\mu_t/\mu$ are set to 5.0\% and 10.0 for the oxidizer stream, and 10.0\% and 100.0 for the fuel stream, as used by Zhu et al. \cite{Zhu2024}.

While this configuration does not exactly replicate how fuel would be introduced into a turbine burner, it is emphasized that this serves as a model to capture the fundamental flow and combustion characteristics. The precise method of fuel introduction is still an active area of research. However, regardless of how the fuel is introduced, the core process always involves two initially separate streams of oxidizer and fuel with a velocity difference, which results in the formation of a reacting mixing layer with a diffusion flame. What makes this configuration more adequate for a turbine burner application and distinguishes it from other studies of reacting mixing layers are the following factors: 1) a large streamwise pressure gradient; 2) high background pressure; 3) elevated oxidizer temperatures; 4) significant freestream turbulence levels; 5) the use of vitiated air as oxidizer; and 6) the flamelet combustion model, which: a) incorporates pressure dependence in the chemical kinetics by adding a fourth dimension to the flamelet tables, and b) accounts for the effects of turbulent fluctuations on sub-grid chemistry using a presumed-shape PDF approach.

\subsection{Numerical Solver}\label{sec:solver}
The solution method employed here is based on the algorithm proposed by Fang et al. \cite{Fang2001}, originally designed for a two-dimensional, steady, laminar, reacting mixing layer. Mehring et al. \cite{Mehring2001}. extended the algorithm for turbulent flows by incorporating the two-equation $k-\omega$ turbulence model. In the present work, the algorithm is modified to include the SST turbulence model and further extended to include sub-grid chemistry modeling via the FPV approach. Fang et al. \cite{Fang2001} proposed a parabolic Crank-Nicholson marching scheme with second-order accuracy in both streamwise and transverse directions. At each streamwise integration step, the governing equations are simultaneously solved at each of the transverse grid points, with freestream values applied as boundary conditions. Mass fractions are then extracted from the flamelet library using values for $\widetilde{Z}$, $\widetilde{Z''^2}$, $\widetilde{C}$ and $\bar{p}$ as lookup parameters. Then, temperature is determined from incremental temperature changes resulting from incremental changes in sensible enthlapy. Marching of the solution requires the specification of inlet conditions for the fuel and oxidizer streams just downstream of the splitter plate. A noteworthy addition in the current work is the use of adaptive mesh refinement in the streamwise direction, ensuring the Courant–Friedrichs–Lewy stability condition for convection and the Von Neumann stability condition for diffusion are both met. More details on the solution method can be found in Refs. \cite{Fang2001,Mehring2001}.

\section{Flow Field Computational Results and Discussion}\label{flow_field_results}
Multiple cases are computed using the turbulence models, oxidizer compositions, and reaction mechanisms discussed previously. The results are presented in four sets: 1) a grid independence study, 2) a reacting mixing layer using the pure air compressible flamelet libraries, 3) the impact of the reaction mechanism's detail on resolved-scale computations, and 4) a reacting mixing layer using the vitiated air compressible libraries. The second set enables direct comparisons with Mehring et al. \cite{Mehring2001} and Zhu et al. \cite{Zhu2024}, who used Westbrook and Dryer's one-step kinetics (OSK) reaction mechanism \cite{WESTBROOK19841}. These comparisons a) highlight flow and flame structure differences resulting from the use of the proposed compressible FPV combustion model compared to the OSK approaches and b) serve as validation of the FPV-boundary-layer approach with comparison to Zhu et al.'s \cite{Zhu2024} higher fidelity two-dimensional Navier-Stokes solutions. 

It should be noted that Zhu et al. \cite{Zhu2024} found that the turbulent mixing layer with initial conditions as specified by Mehring et al. \cite{Mehring2001} might not ignite. To address this, the authors increased the pre-exponential frequency factor of the global reaction by an order of magnitude compared to the value used by Mehring et al. \cite{Mehring2001} and given by Westbrook and Dryer \cite{WESTBROOK19841}. The present study similarly observes that the flow does not ignite using the FPV approach. To maintain consistency with the conditions employed by Zhu et al., the mean production rate of the progress variable $\widetilde{\dot{\omega}}_C$ is also increased by an order of magnitude. Although not reported here, cases with lower pressure gradient were also tested and naturally occurring ignition, i.e., without increasing $\widetilde{\dot{\omega}}_C$, was found to occur at favorable streamwise pressure gradients of 50 atm/m or lower.

The results presented in this section are generated using either the $k-\omega$ or SST turbulence models with the corresponding freestream turbulence parameters specified in Section \ref{sec:flow_conf}, allowing for fair comparisons with the reference studies by Mehring et al. \cite{Mehring2001} and Zhu et al. \cite{Zhu2024}. It is important to note that all comparisons are made between results that use the same turbulence model and parameters. 

Unless otherwise specified, all flow variables reported within this section correspond to resolved-scale Favre averages.

\subsection{Grid Independence}
To assess grid independence for the current code, which now employs the compressible FPV formulation as a novelty compared to previous iterations \cite{Fang2001, Mehring2001}, three cases are computed using the SST turbulence model and the reduced FFCM-13 reaction mechanism. These cases differ only in their mesh resolution: a coarse mesh with 4,960,000 computational cells, a fine mesh with 38,928,000 cells, and an extra-fine mesh with 55,323,000 cells serving as a baseline for error computation.

\begin{figure}
\centering
\begin{subfigure}{.5\textwidth}
  \centering
  \includegraphics[width=\linewidth]{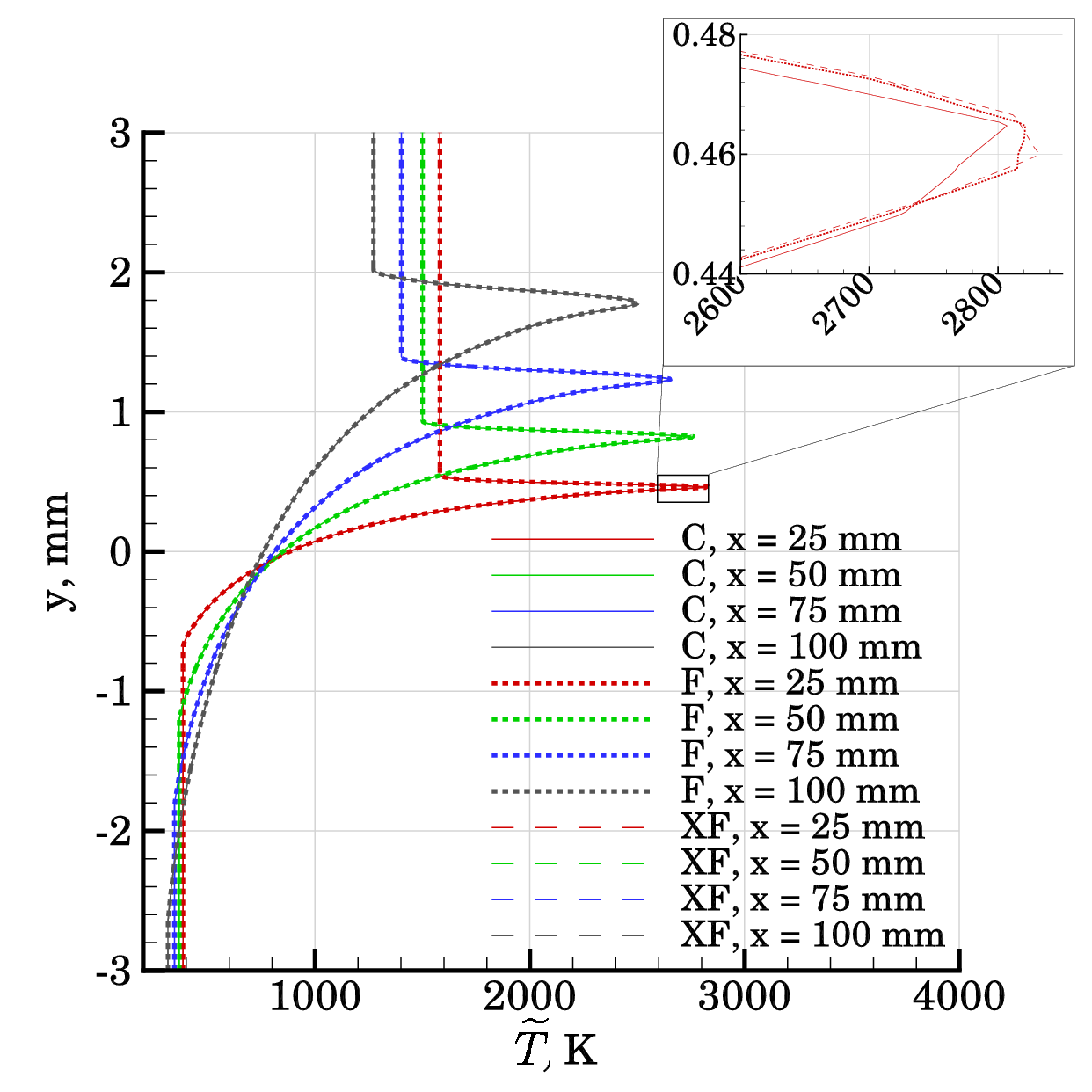}
  \caption{Temperature at four streamwise locations.}
  \label{fig:temp_grid_profiles}
\end{subfigure}%
\begin{subfigure}{.5\textwidth}
  \centering
  \includegraphics[width=3.25in]{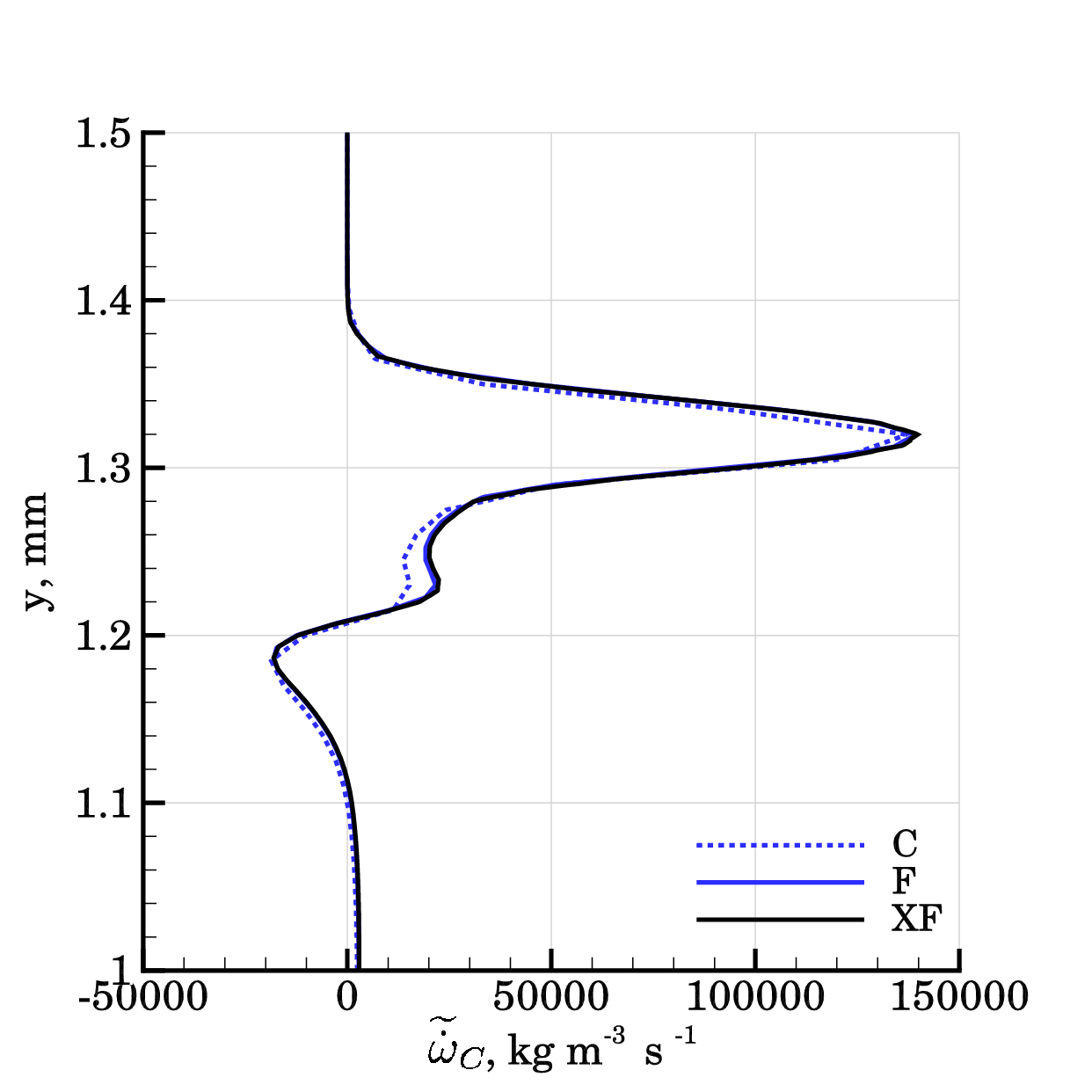}
  \caption{Production rate of the progress variable $\widetilde{\dot{\omega}}_C$ at x = 75 mm.}
  \label{fig:omC_grid_profiles}
\end{subfigure}
\caption{Temperature and progress variable production rates for varying mesh sizes. C: coarse; F: fine; XF: extra-fine.}
\label{grid_ind}
\end{figure}

Figure \ref{fig:temp_grid_profiles} presents temperature profiles at four streamwise locations (x = 25, 50, 75, and 100 mm) for these three cases. The solutions are practically indistinguishable across all three grid levels. Peak temperatures deviate by less than 2\% from the peak temperature predicted by the extra-fine case, with the coarse grid reporting the lowest value.
Although resolved-scale temperature has been shown to be grid-independent, we seek validation of the production rate of the resolved-scale progress variable $\widetilde{\dot{\omega}}_C$, a parameter extracted from the flamelet libraries and one of the stiffest variables in this study. Figure \ref{fig:omC_grid_profiles} reports $\widetilde{\dot{\omega}}_C$ at the streamwise location of x = 75 mm. The coarse grid solution reports values noticeably lower than those for the fine grid solutions, resulting in an overall lower heat-release rate and marginally lower peak temperatures. The near-identical results of the fine and extra-fine cases indicate that grid independence has been achieved, suggesting that these grid resolutions are sufficient to correctly capture $\widetilde{\dot{\omega}}_C$, the driving term of the combustion process.

\begin{figure}
    \centering
    \includegraphics[]{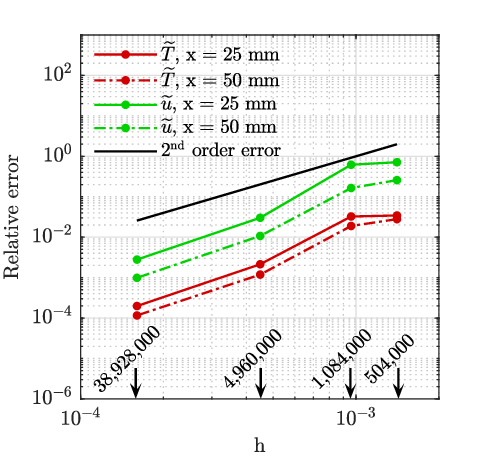}
    \caption{Relative error of temperature and streamwise velocity component as a function of grid spacing h.}
    \label{fig:grid_error}
\end{figure}

Further confidence is attained by evaluating the change of relative errors as function of grid size, which are provided in Fig. \ref{fig:grid_error}. Here, the errors for resolved-scale temperature and streamwise velocity component are computed as the relative difference of mass-averaged field variables over a streamwise cross-section with respect to those in the extra-fine case. The grid spacing is determined as $h=\sqrt{1/N_c}$, where $N_c$ represents the total number of computational cells for each case. The relative errors of all resolved-scale variables, including those not shown, decrease with decreasing grid spacing at a rate corresponding to second-order accuracy. This second-order accuracy rate agrees with the spatial discretization accuracy of the numerical solver. 

\subsection{Pure Air Turbulent Mixing Layer}

\begin{figure}
    \centering
    \includegraphics[width=6.5in]{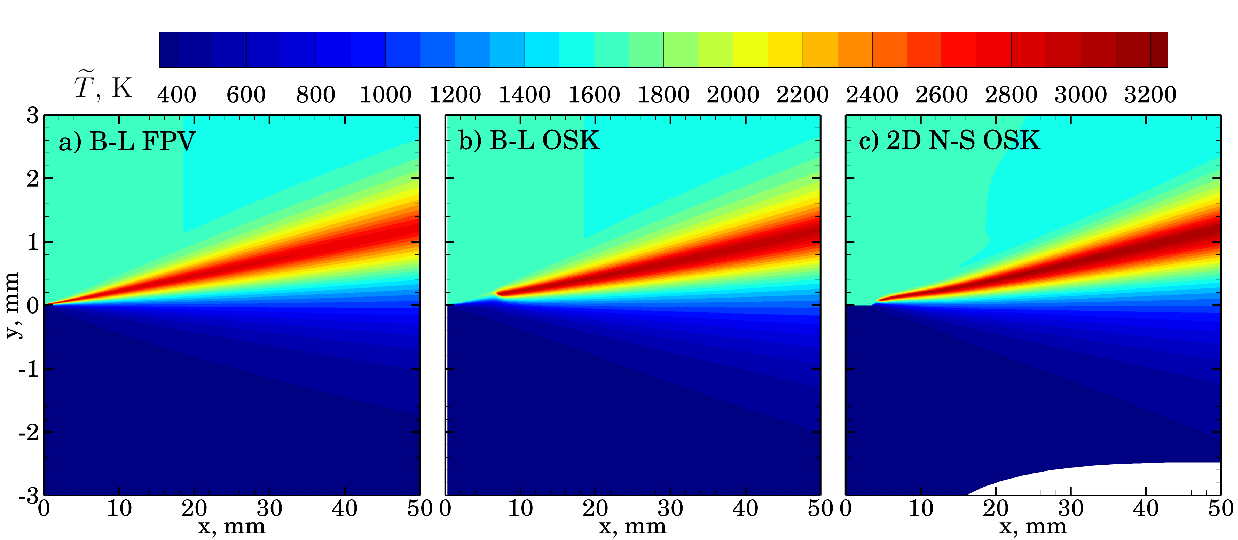}
    \caption{Temperature contours for the pure air cases. a) present; b) Mehring et al. \cite{Mehring2001}; c) Zhu et al. \cite{Zhu2024}.}
    \label{fig:tcont}
\end{figure}

The following presents computational results for a reacting mixing layer using pure air in the oxidizer stream and the reduced FFCM-13 reaction mechanism. These results are obtained using the $k-\omega$ turbulence model to enable direct comparison with the studies by Mehring et al. \cite{Mehring2001} and Zhu et al. \cite{Zhu2024}. Note that Zhu et al. \cite{Zhu2024} also performed cases using the $k-\omega$ model for comparison with the results of Mehring et al. \cite{Mehring2001}.  

When the incoming streams of hot air and fuel vapor meet at x = 0 mm, they start to mix, forming thermal and velocity mixing layers. The flow ignites shortly after the two streams meet, establishing a diffusion flame within the mixing layer. The flame, represented by the high-temperature regions observed in Fig. \ref{fig:tcont}, is located where reactant concentrations are at their stoichiometric values, $Z_{st}$ = 0.055 for the methane-pure air reaction. Figures \ref{fig:tcont}b and \ref{fig:tcont}c plot temperature contours for the OSK boundary-layer (B-L OSK) results by Mehring et al. \cite{Mehring2001} and the OSK two-dimensional Navier-Stokes (2D N-S OSK) solution by Zhu et al., respectively. Noticeably, all results show different ignition delays. The present computations report the lowest ignition delay at x = 0.5 mm, followed by Zhu et al. \cite{Zhu2024} at around x = 5 mm, and finally Mehring et al. \cite{Mehring2001} at about x = 10 mm.

The discrepancy in ignition delay for the OSK studies was explained by Zhu et al. \cite{Zhu2024} to be the result of: 1) the pressure gradient in Zhu et al. \cite{Zhu2024} being enforced using a convergent-divergent nozzle (observed as the white region in the lower half of Fig. \ref{fig:tcont}c) which cannot exactly replicate the prescribed boundary-layer approximation value near the  inlet where the flow is strongly two-dimensional in nature; 2) the boundary-layer approximation being not sufficiently accurate in this initial highly two-dimensional flow region. On the other hand, the shorter ignition delay predicted by the present results compared with the OSK studies is attributed to FPV approach, wich has the more elaborate combustion model, with much faster chemistry than the OSK counterparts.

When comparing B-L solutions to the 2D N-S, differences in mixing layer growths are observed. Both B-L results show nearly linearly developing flames throughout the whole domain, while the 2D N-S presents vertical flame deflection towards the air-side, near the inlet. This is explained again by the lack of accuracy of the B-L approach in the region immediately downstream of the edge of the computational domain, near x = 0 mm.

\begin{figure}
\centering
\begin{subfigure}{.5\textwidth}
  \centering
  \includegraphics[width=3.25in]{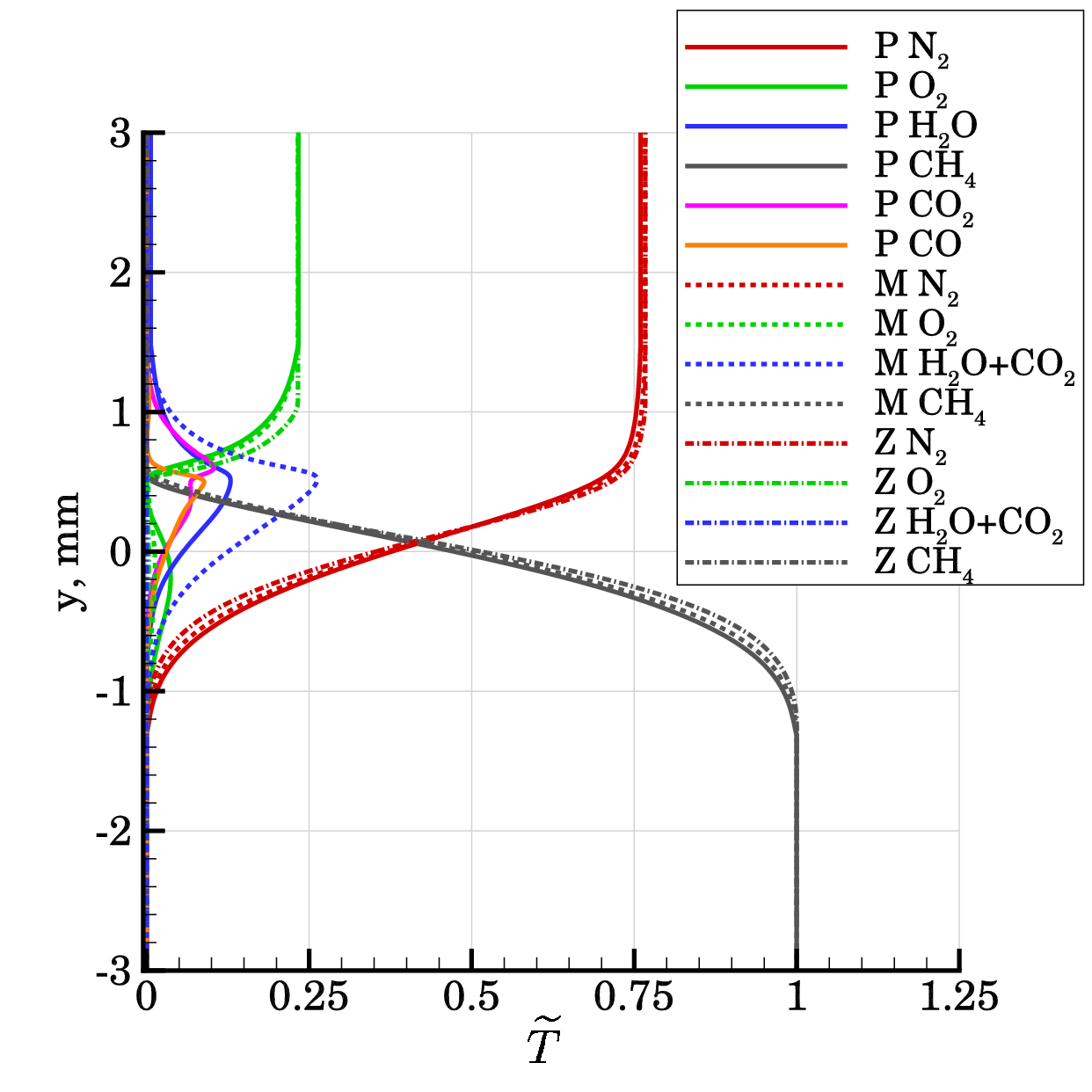}
  \caption{Major species.}
  \label{fig:y_major}
\end{subfigure}%
\begin{subfigure}{.5\textwidth}
  \centering
  \includegraphics[width=3.25in]{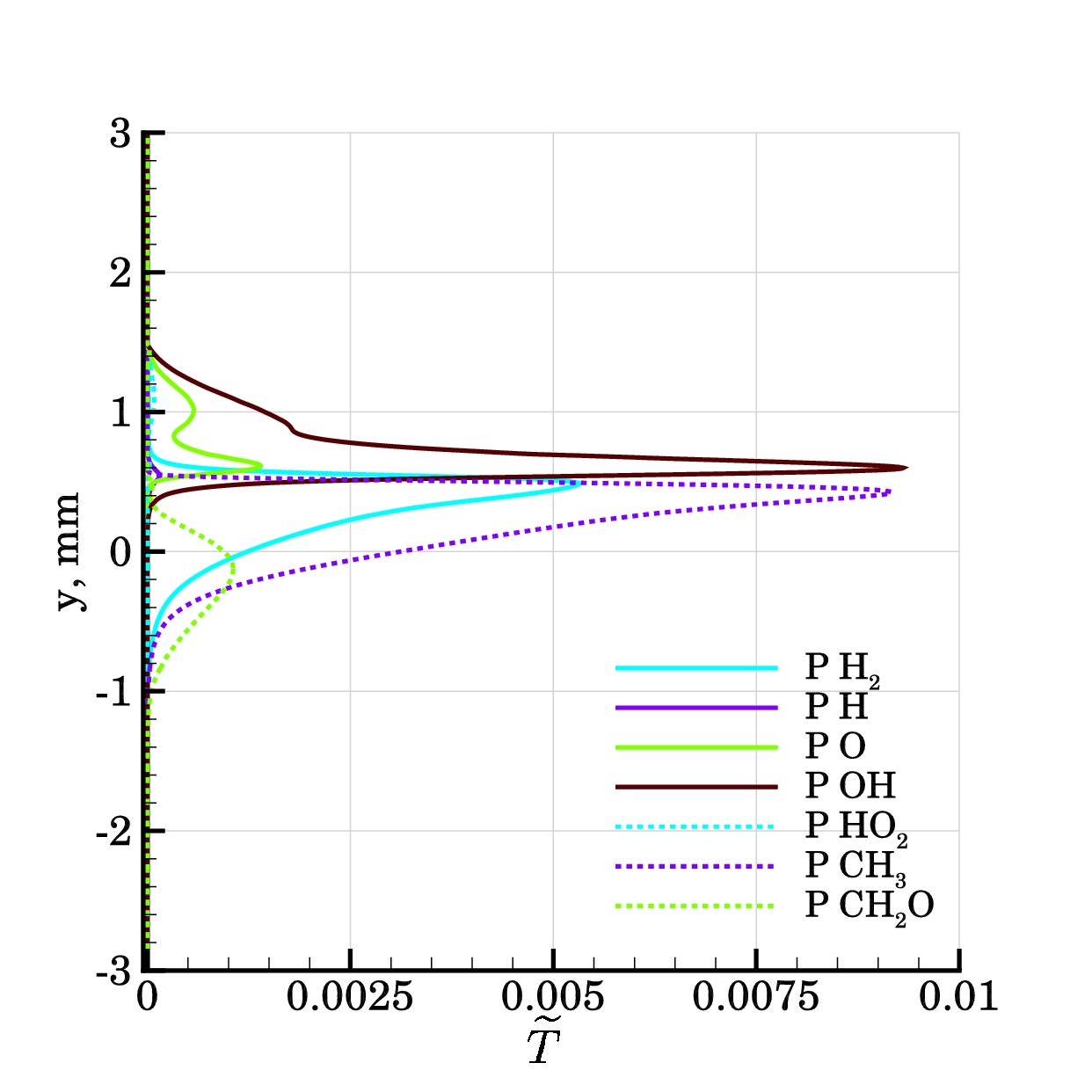}
  \caption{Minor species.}
  \label{fig:y_minor}
\end{subfigure}
\caption{Species mass fraction profiles at x = 25 mm for the pure air cases. P: present; M: Mehring et al. \cite{Mehring2001}; Z: Zhu et al. \cite{Zhu2024}.}
\label{fig:y_2r}
\end{figure}

Figures \ref{fig:y_major} and \ref{fig:y_minor} show major and minor species mass fractions profiles at a constant streamwise location of x = 25 mm, respectivley. The profiles pertaining to the present FPV approach are constructed from sub-grid mass fractions extracted from the locus of flamelets spanning the transverse coordinate, one per each computational cell. Each one of these flamelets is subjected to different values of the resolved-scale progress variable and, thus, represented by a different flamelet solution along the "S-shaped" curve (see Fig. \ref{fig:s_shaped_pure_air}). Furthermore, each flamelet is subjected to a different level of turbulence determined by the distribution of $\widetilde{Z''^2}$. Mass fractions of reactants ($\mathrm{O_2}$ and $\mathrm{CH_4}$) and $\mathrm{N_2}$ show better agreement between the two B-L solutions compared to the 2D N-S solutions, stemming from the differences in mixing layer growths reported above. Nitrogen is inert and simply diffuses from the air-side to the fuel-side. The discrepancy between the freestream values of $\mathrm{N_2}$ is attributed to the fact that Mehring et al. \cite{Mehring2001} and Zhu et al. \cite{Zhu2024} did not include $\mathrm{H_2O}$ in their air-streams. In the present work, $\mathrm{H_2O}$ is included in the air streams to better reflect realistic atmospheric conditions. That said, the concentration of $\mathrm{H_2O}$ in the freestream is minimal (<1\% by mass) and does not affect the oxidizer ($\mathrm{O_2}$) concentration relative to the other studies, as it simply replaces ($\mathrm{N_2}$). Therefore, the inclusion of $\mathrm{H_2O}$ in the freestream is not expected to have a significant impact on combustion.

Product mass fractions peak in the reaction zone, which is located near $Z_{st}$. $\mathrm{H_2}$, $\mathrm{H}$, $\mathrm{O}$, $\mathrm{OH}$, $\mathrm{HO_2}$, $\mathrm{CH_3}$ and $\mathrm{CH_2O}$ are all minor species with a maximum combined mass fraction amounting to roughly 1.5\%. Noticeable differences are observed between the OSK and FPV results in terms of major product mass fractions, namely $\mathrm{H_2O}$ and $\mathrm{CO_2}$, which can be explained by the significant distinctions in reaction pathways considered by both combustion models.

\begin{figure}
    \centering
    \includegraphics[width=6.5in]{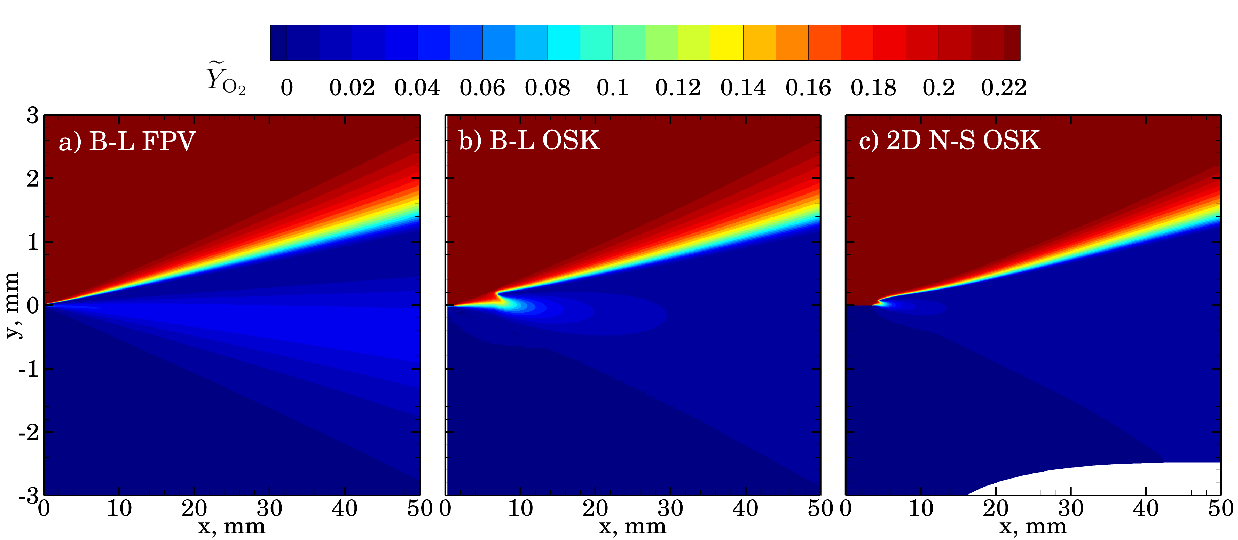}
    \caption{Oxygen mass fraction contours for for the pure air cases. a) present; b) Mehring et al. \cite{Mehring2001}; c) Zhu et al. \cite{Zhu2024}.}
    \label{fig:o2cont}
\end{figure}

Results from all studies indicate that entrainment of $\mathrm{O_2}$ into the fuel-side occurs before iginition, leading to local $\mathrm{O_2}$ maxima located below the center-line. This phenomenon is more clearly observed in the $\mathrm{O_2}$ mass fraction contour plots reported in Fig. \ref{fig:o2cont}. Zhu et al. \cite{Zhu2024} explained this as a consequence of the flame stand-off from the edge of the splitter plate. The region upstream of the ignition point is dominated by convection and diffusion, the latter allowing for oxygen to be "trapped" on the fuel-side. Given that the resulting mixture composition is beyond the flammability limit, this entrained oxygen does not react and simply diffuses downstream. When comparing the two OSK model results, more $\mathrm{O_2}$ is entrained in Mehring et al. \cite{Mehring2001} due to the larger ignition delay. The FPV model shows inconsistencies pertaining to the lack of diffusion of $\mathrm{O_2}$ entrained on the fuel-side and noticeably higher local $\mathrm{O_2}$ maxima in the fuel-side despite the shortest flame stand-off distance. This is explained (at least in part) by the fact, that in the FPV model, diffusion does not act directly on the species gradients but instead acts indirectly through the gradients of $\widetilde{Z}$, $\widetilde{Z''^2}$ and $\widetilde{C}$. This represents a significant shortcoming of the FPV approach, as resolved-scale turbulent diffusion is not fully experienced by the sub-grid flamelets, even when considering the $\beta$-PDF convolution process.

\begin{figure}
\centering
\begin{subfigure}{.5\textwidth}
  \centering
  \includegraphics[width=3.25in]{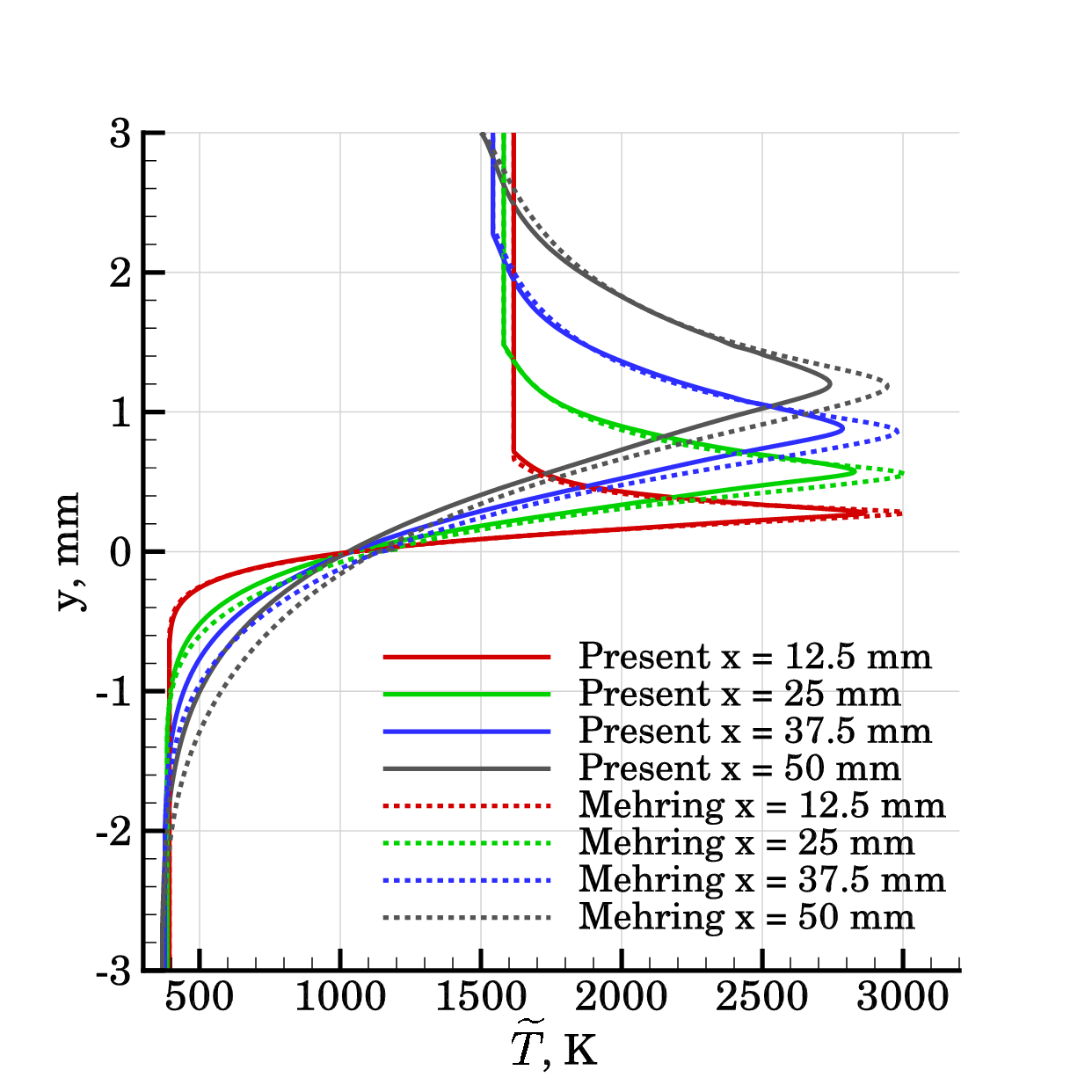}
  \caption{Present (B-L FPV) and Mehring et al. (B-L OSK) \cite{Mehring2001}.}
  \label{fig:t_mehring_present_k-w}
\end{subfigure}%
\begin{subfigure}{.5\textwidth}
  \centering
  \includegraphics[width=\linewidth]{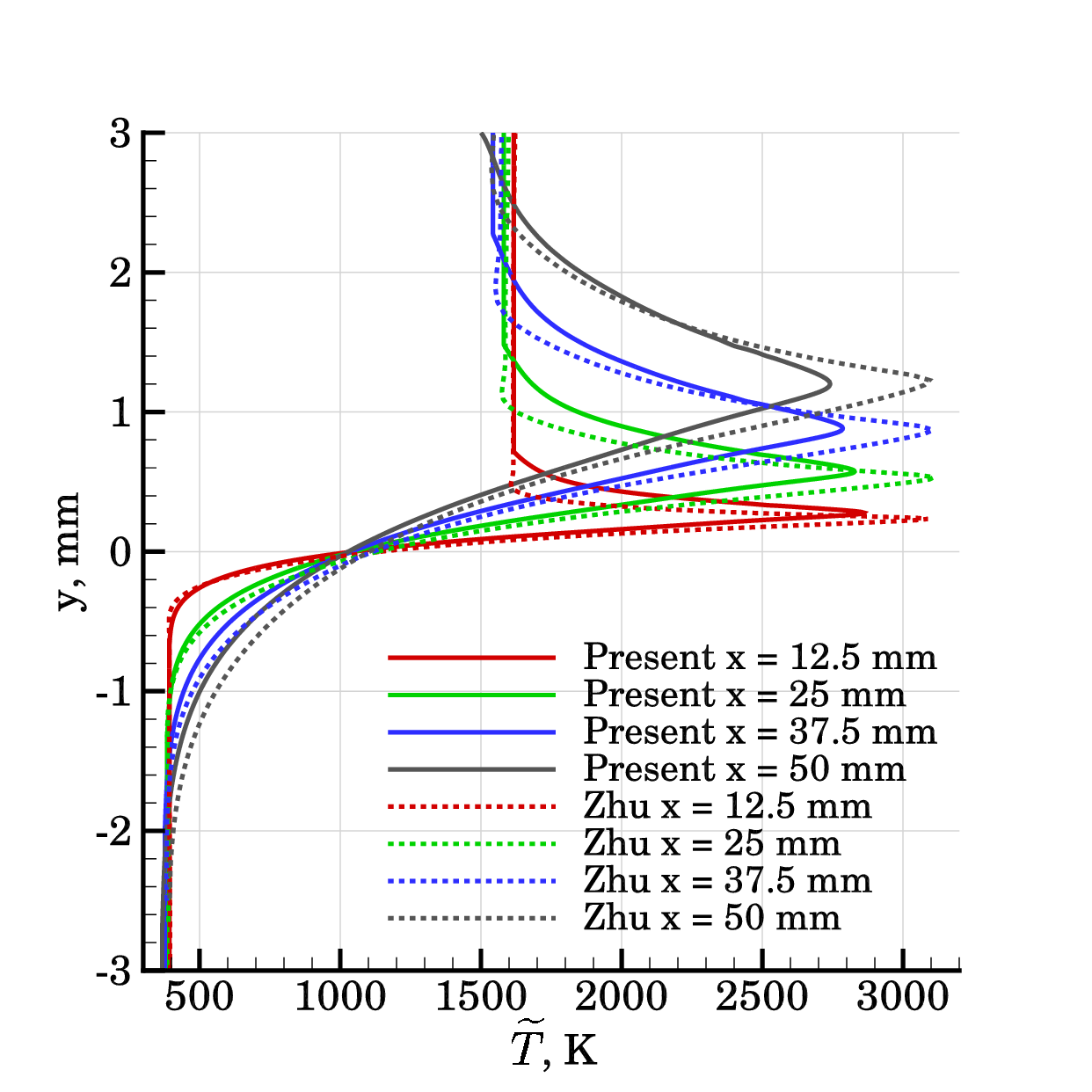}
  \caption{Present (B-L FPV) and Zhu et al. (2D N-S OSK) \cite{Zhu2024}.}
  \label{fig:t_zhu_present_k-w}
\end{subfigure}
\caption{Comparison of temperature profiles at different streamwise locations for the pure air cases.}
\label{fig:t_profiles_literature}
\end{figure}

Figure \ref{fig:t_mehring_present_k-w} illustrates temperature profiles of the present work together with those of Mehring et al. \cite{Mehring2001} at four different streamwise locations (x = 12.5, 25, 37.5 and 50 mm). freestream values are in good agreement; however, peak temperatures obtained by the FPV model are significantly lower, with a maximum absolute difference of around 200 K. This reduction is attributed to three factors: 1) loss of heat to dissociation and radical formation, which is not present in the OSK chemistry model; 2) lower peak mixture-averaged specific heat capacity $c_p$ due to the significant amounts of CO in the FPV model solution; and 3) the FPV approach accounts for turbulent interaction with the chemistry through the reduction of peak temperatures as a consequence of increasing reaction zone thickness in the flamelet geometry. In contrast, the OSK model neglects the influence of turbulence on the reaction rate except for the use of the turbulent fuel and air mass fractions in the global rate equation. Peak temperatures in both results shown in Fig. \ref{fig:t_mehring_present_k-w} decrease along the flow direction due to flow acceleration inside and outside the mixing layer. However, the rate of reduction is slightly greater for the present results. This can be explained by the pressure sensitivity of the FPV approach, in which sub-scale net reaction rates decrease with decreasing background pressure.  

Setting peak temperature aside, the present temperature profiles agree well with those of Mehring et al. \cite{Mehring2001} at x = 12.5 mm, but are slightly biased towards the air-side as we progress downstream. This behavior can be attributed at least in part to the lack of diffusion in the entrained oxygen pertaining to the FPV results. 

\begin{figure}
\centering
\begin{subfigure}{.5\textwidth}
  \centering
  \includegraphics[width=3.25in]{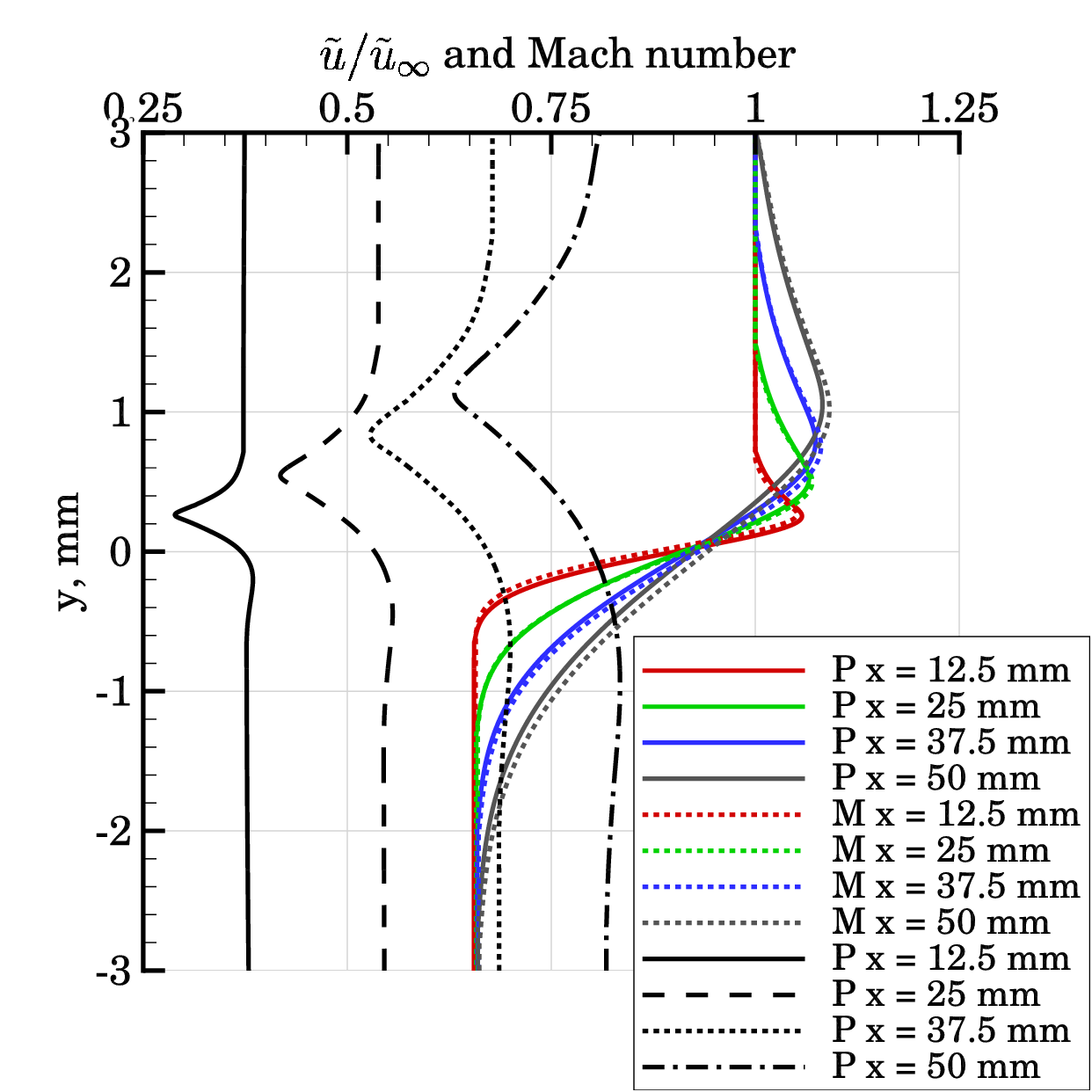}
  \caption{Velocity (colored) and Mach number (black).}
  \label{fig:u_mehring_present_k-w}
\end{subfigure}%
\begin{subfigure}{.5\textwidth}
  \centering
  \includegraphics[width=3.25in]{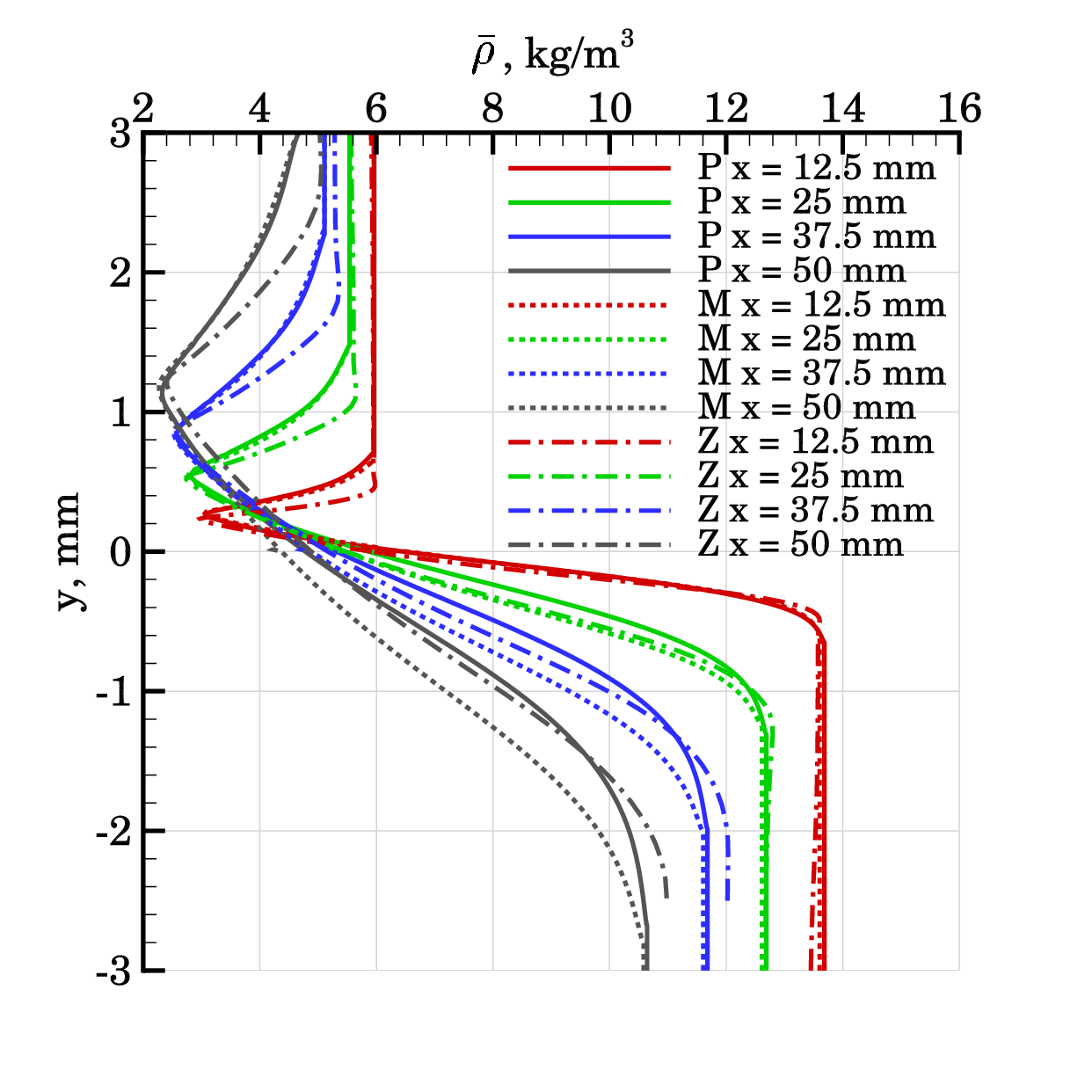}
  \caption{Density.}
  \label{fig:rho_zhu_present_k-w}
\end{subfigure}
\caption{Comparison of flow variable profiles at different streamwise locations. P: present; M: Mehring et al.; \cite{Mehring2001}; Z: Zhu et al. \cite{Zhu2024}.}
\label{fig:flow_variables_k-w}
\end{figure}

With constant pressure along the transverse direction, the density dips in the flame region as a result of higher temperature. Consequently, the lighter gas in this region gets accelerated more, resulting in a peak in velocity. This is illustrated in Fig. \ref{fig:flow_variables_k-w} which shows normalized streamwise velocity, Mach number, and density at four streamwise locations. Here, the velocity is normalized by the air-side freestream value at each streamwise location. Good agreement with the results of Mehring et al. \cite{Mehring2001} is observed on the air-side. As for the fuel-side, the FPV model yields a thinner mixing layer stemming from the lower temperature predictions in this region. The results by Mehring et al. \cite{Mehring2001} show slightly lower density troughs and slightly higher velocity peaks, consistent with the higher peak temperatures in their study. The Mach number dips within the mixing layer and reaches a minima in the flame zone where the temperature peaks. Nevertheless, the freestream Mach number values clearly indicate that a low-Mach number approximation is not valid and thus, justify the compressible modifications to the FPV formulation presented in Section \ref{sec:comb_model}.

Figures \ref{fig:t_zhu_present_k-w} and \ref{fig:rho_zhu_present_k-w} show temperature and density profiles obtained by the FPV approach together with those from Zhu et al. \cite{Zhu2024}. Overall, the respective results agree quite well and show the same behaviors described in the discussion with respect to the results of Mehring et al. \cite{Mehring2001}. However, two major differences are observed: 1) both boundary-layer solutions are more diffusive compared to the two-dimensional Navier-Stokes solutions as indicated by their wider density and temperature mixing layers resulting from stronger turbulent diffusion; and 2) the two boundary-layer solutions show lower peak temperatures. Zhu et al. \cite{Zhu2024} argued this to be a consequence of the stronger turbulent diffusion which induces lower temperature within the flame.

Although not reported here, the pure air results just shown were also generated using the SST model with higher freestream turbulence intensities. Despite differences in mixing layer growth and flame thickness, combustion characteristics such as peak temperatures and species production were found to remain largely similar to the results using the $k-\omega$ model with lower freestream turbulence levels. The key trends distinguishing the FPV combustion model from the OSK model are consistent across both turbulence models.

\subsection{Impact of Reaction Mechanism Detail}\label{rs_mech_comp}
Two additional computational cases are conducted to compare the resolved scale results of two flamelet models with different levels of detail in their reaction mechanisms. In these cases, pure air constitutes the oxidizer stream, and the SST turbulence model is utilized. All parameters remain consistent with those described in Section \ref{sec:flow_conf}, with exception for the incoming stream velocities. The velocity difference between the two streams is increased fivefold, resulting in initial velocities of 22.5 m/s and 225.0 m/s for the fuel and oxidizer streams, respectively. This adjustment is motivated by the findings depicted in Fig. \ref{fig:ffcm_vs_ffcm13}a), where the flamelet model associated with the FFCM-1 reaction mechanism demonstrates greater resilience to higher strains before extinguishing. Consequently, augmenting the resolved-scale strain rate is intended to accentuate distinctions in ignition behavior between the resolved-scale results produced by both mechanisms.

\begin{figure}
\centering
\begin{subfigure}{.5\textwidth}
  \centering
  \includegraphics[width=3.25in]{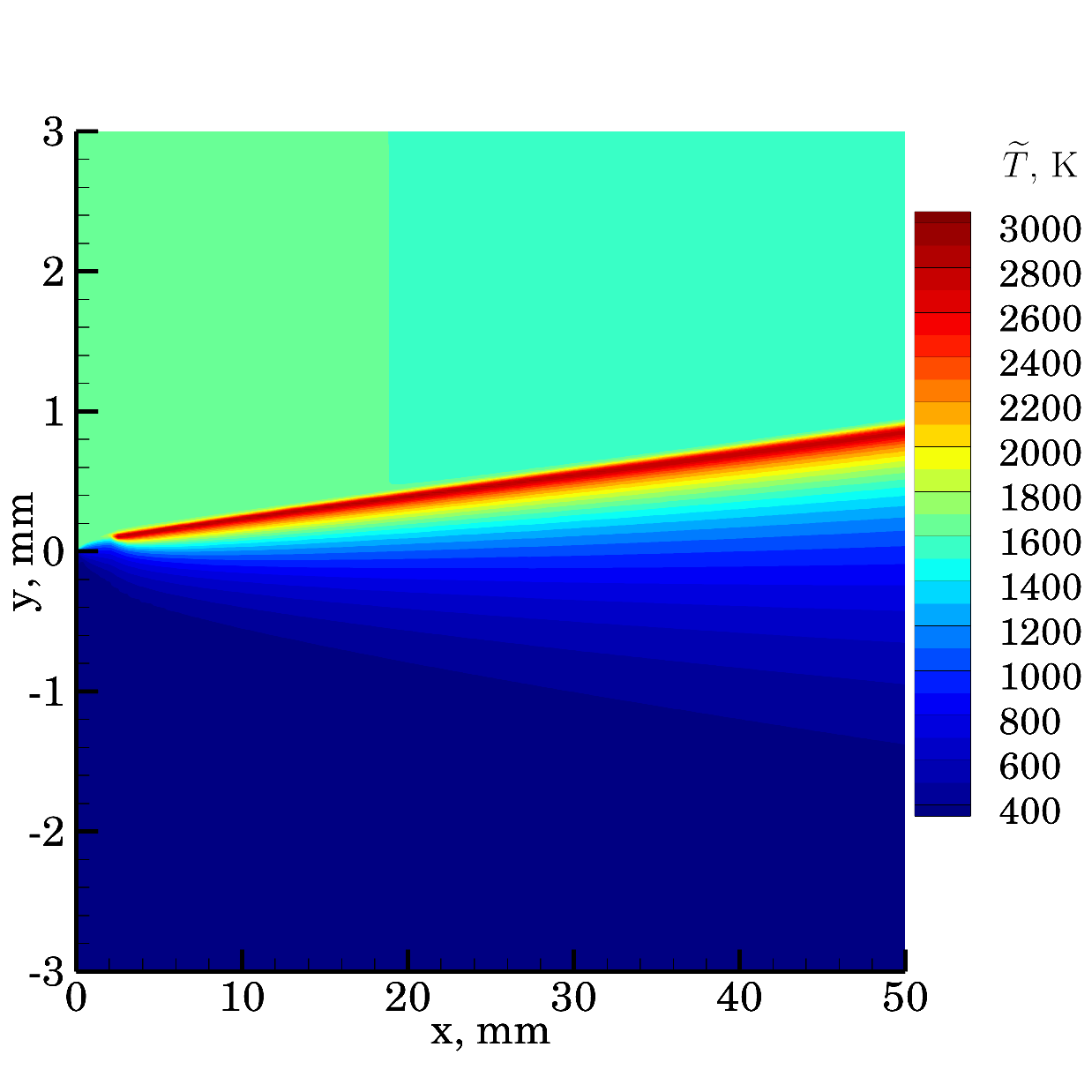}
  \caption{FFCM-13.}
  \label{fig:ffcm13_t}
\end{subfigure}%
\begin{subfigure}{.5\textwidth}
  \centering
  \includegraphics[width=3.25in]{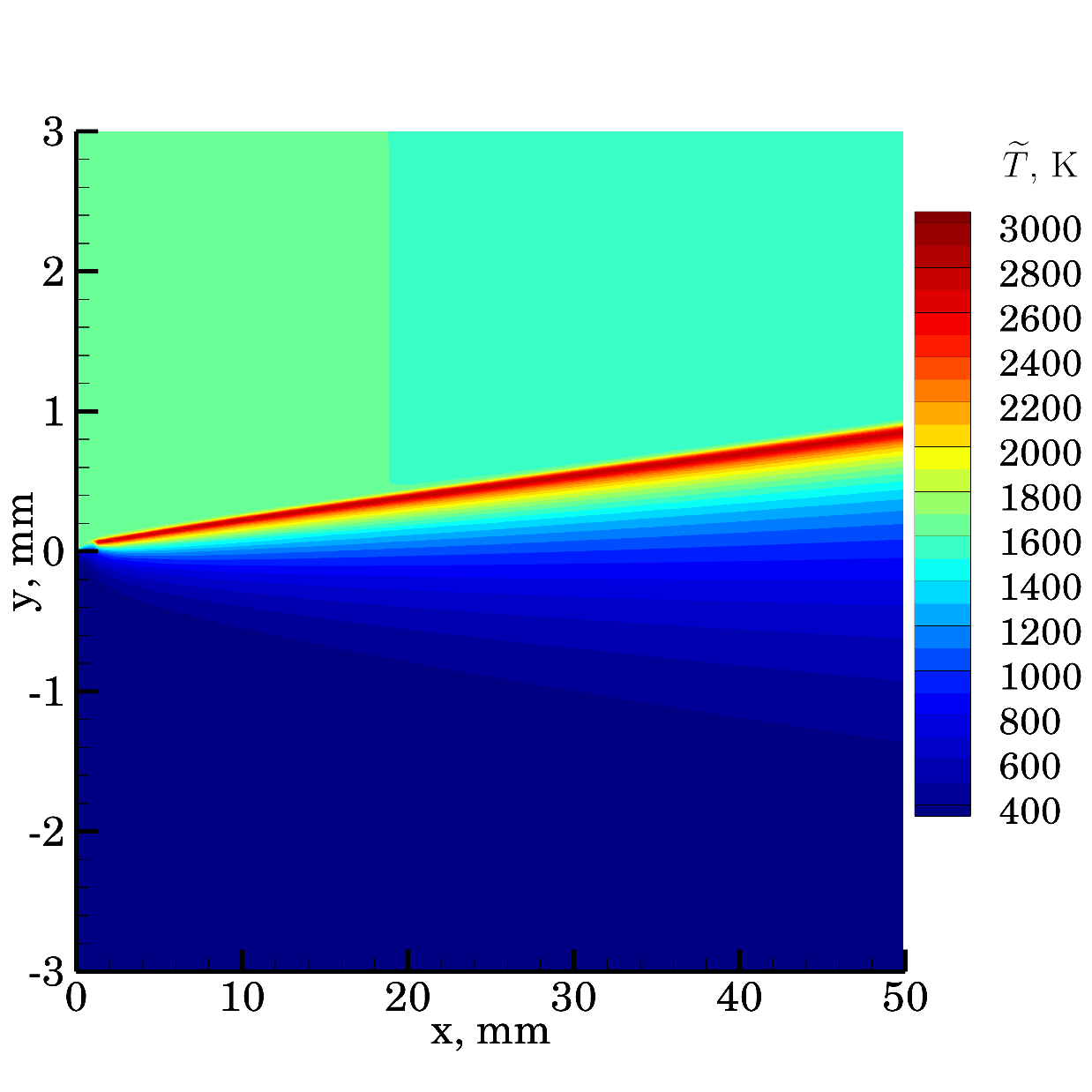}
  \caption{FFCM-1.}
  \label{fig:ffcm1_t}
\end{subfigure}
\caption{Temperature contours for pure air combustion obtained from both the reduced FFCM-13 and full FFCM-1 reaction mechanisms.}
\label{fig:mechanism_comp}
\end{figure}

Figure \ref{fig:mechanism_comp} illustrates mean temperature contours obtained using the skeletal reduction FFCM-13 and the full FFCM-1 reaction mechanisms. With the heightened velocity difference between the incoming streams, notable disparities in mixing layer growth compared to previous computational cases are anticipated. However, the focus here lies on studying how differences between the two mechanisms arising at the sub-grid level (presented in Section \ref{sec:ffcm1_ffcm13_flamelet}) affect resolved-scale computations. Notably, distinct flame standoffs are observed, with the FFCM-1 exhibiting earlier ignition than the FFCM-13 mechanism. 

At the sub-grid level, flamelets are governed by the strain rate, meaning different flame states (flamelets) are described for varying values of $\chi_{st}$ (see Figs. \ref{fig:s_vit_air} and \ref{fig:s_shaped_new}). The strain rate is the controlling physical parameter in this context so we initially aim to explain the difference in flame standoffs in terms of the resolved-scale strain rate. At the resolved scale before ignition takes place, due to the velocity difference between the incoming reactant streams, the highest resolved-scale strain rate occurs where the two flows meet (at x = 0). As one moves downstream at the same y location, the resolved-scale strain rate decreases due to the diffusion of velocity gradients. The earlier ignition of the FFCM-1 mechanism could thus be attributed to its ability to tolerate higher strain rates (as shown in Fig. \ref{fig:ffcm_vs_ffcm13}a).

However, while this explanation appears coherent with the results, it is not valid due to the use of the FPV approach. Once the sub-grid flamelet solutions are mapped to the progress variable, the connection between the flamelet and resolved-scale strain rate is lost. There is no direct relationship between the resolved-scale mean progress variable value $\widetilde{C}$ and the resolved-scale strain rate magnitude. As a result, the faster ignition of the FFCM-1 mechanism at the resolved scale is not directly due to its higher tolerance to strain rate, but rather to the increased production rate of the progress variable, as shown in Figs. \ref{fig:ffcm_vs_ffcm13}b-c).

\begin{figure}
\centering
\begin{subfigure}{.5\textwidth}
  \centering
  \includegraphics[width=3.25in]{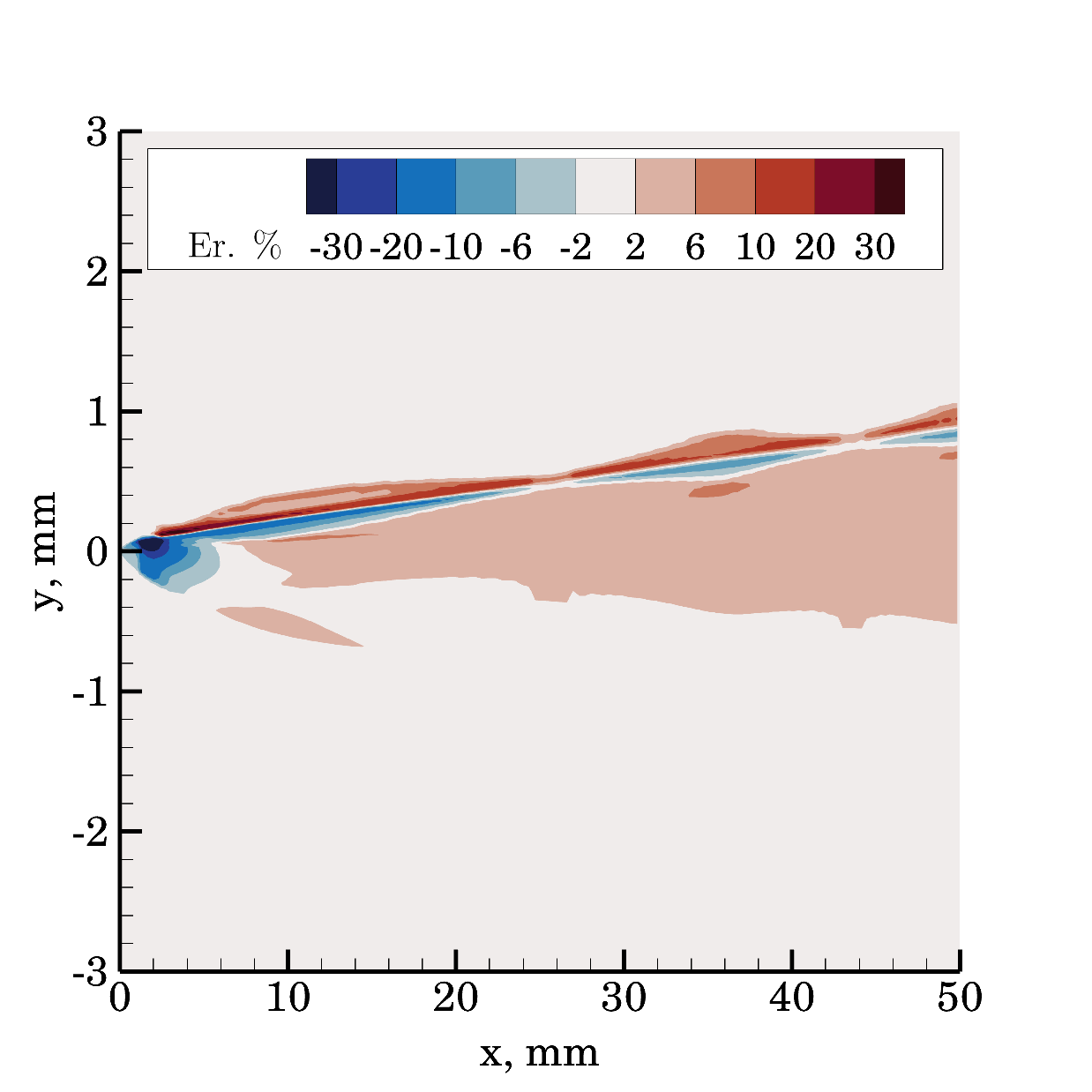}
  \caption{Relative temperature error (FFCM-13 w.r.t. FFCM-1).}
  \label{fig:t_rel_error}
\end{subfigure}%
\begin{subfigure}{.5\textwidth}
  \centering
  \includegraphics[width=3.25in]{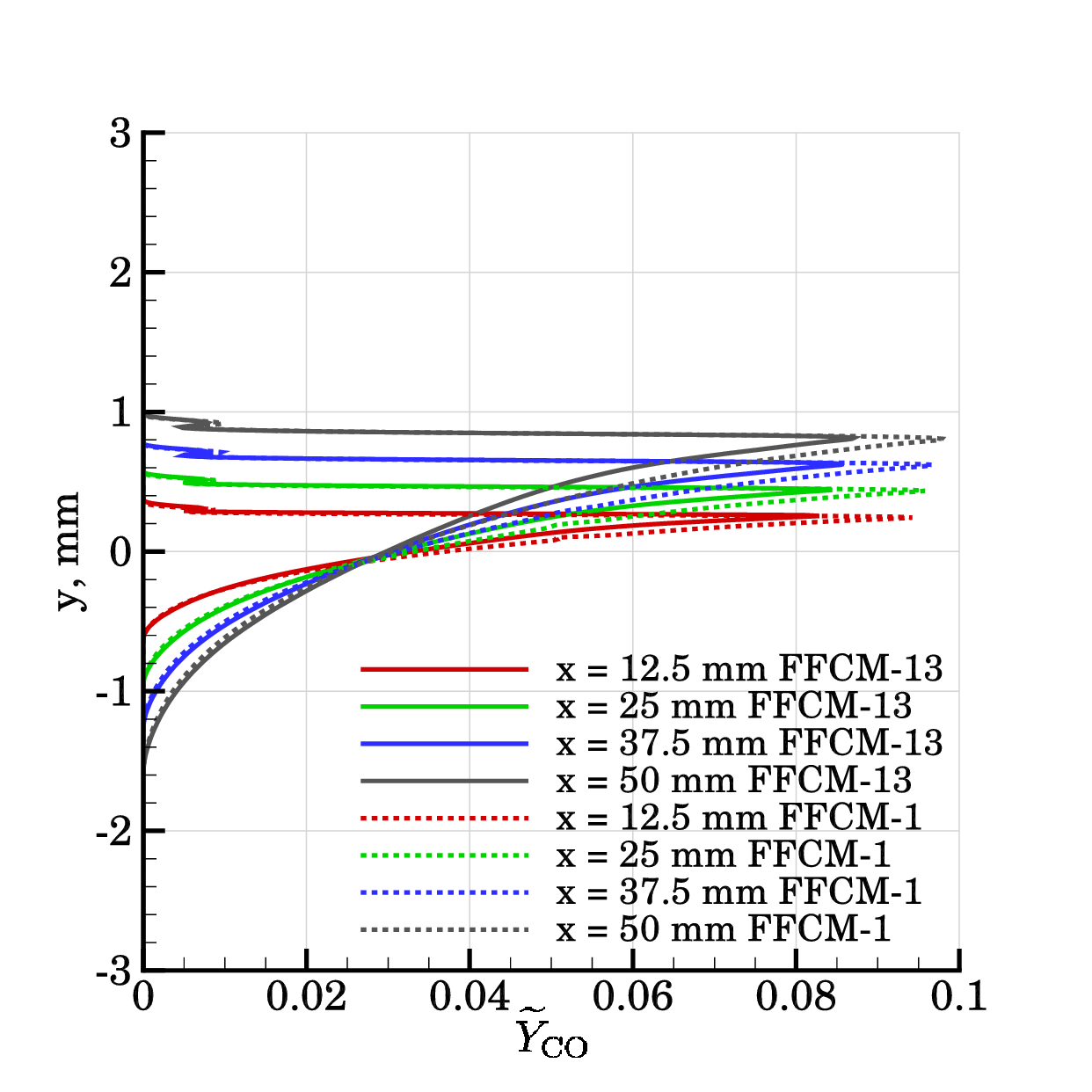}
  \caption{CO mass fraction profiles.}
  \label{fig:co_comp}
\end{subfigure}
\caption{Relative temperature error and CO mass fractions.}
\label{fig:mech_error}
\end{figure}

This finding is significant because it highlights a disconnect between sub-scale strain rate and resolved-scale mechanics when using the FPV approach. Although the flamelet model uses sub-grid strain rate as the controlling physical parameter, it is found that the resolved-scale combustion process is instead driven by chemistry (via the progress variable) rather than strain rate dynamics. This means that based on the results in Figs. \ref{fig:ffcm_vs_ffcm13}b-c), The FFCM-1 reaction mechanism will always ignite sooner than the FFCM-13 mechanism, regardless of the resolved-scale strain rates. Even if the resolved-scale strain rate is eliminated before ignitions by removing the velocity difference between the two reactant streams and not considering a pressure gradient, the FFCM-1 mechanism will still ignite sooner. 

Figure \ref{fig:t_rel_error} plots the relative error in temperature obtained with the FFCM-13 mechanism with respect to that obtained with the FFCM-1 mechanism. The dark blue region just downstream of x = 0.0 mm represents the difference in flame stand-off between both reaction mechanisms. Subsequently, the dark red and blue regions in the flame zone suggests that the retarded ignition of the FFCM-13 mechanism shifts the flame location slightly more toward the oxidizer side. Throughout the entirety of the mixing layer, the FFCM-13 mechanism tends to slightly over-predict temperature (light red zone corresponding to less than 6\% relative error), agreeing with the findings of Section \ref{sec:ffcm1_ffcm13_flamelet}. Given the differing levels of chemical kinetic detail between the two reaction mechanisms, variations in product and radical composition are evident and expected. Figure \ref{fig:co_comp} presents profiles of CO mass fraction at four distinct streamwise locations, indicating that the FFCM-1 mechanism yields approximately 10\% more CO in the flame zone, which is indicative of it having faster chemistry than the FFCM-13 mechanism.

These findings clearly indicate that the differences observed in the sub-scale flamelet solutions carry over to the resolved-scale, with predictions of flame properties including ignition, temperature, and species production differing among the two mechanisms. While experimental validation is currently lacking, we can assert that, assuming the reliability of the FFCM-1 mechanism, employing this more detailed mechanism, which provides a more comprehensive representation of chemical kinetics, should offer greater accuracy in capturing the combustion process. However, if the differences highlighted here are deemed negligible, there is a compelling case for utilizing the reduced FFCM-13 mechanism, as it offers approximately a two-thirds reduction in both storage requirements and computational time.

\subsection{Vitiated Air Turbulent Mixing Layer}\label{sec:vit_air_res}
In this case, the oxidizer stream consists of vitiated air with a composition corresponding to pre-burning in the primary upstream combustion chamber. This configuration aims to better replicate the flow conditions expected within the turbine-burner. The reduced FFCM-13 mechanism is used alongside the SST turbulence model, with the latter enabling a direct comparison with the results by Zhu et al. \cite{Zhu2024}. Although using the full reaction mechanism would provide a more detailed representation of the chemical kinetics, the temperature predictions between the two mechanisms in the reaction zone differ by less than 6\%, as shown in the previous section. Additionally, the use of the full mechanism complicates the generation of flamelet tables. Therefore, the FFCM-13 mechanism is considered sufficiently adequate for this case. All other flow conditions are kept identical to those in the pure air cases.
\begin{figure}
\centering
\begin{subfigure}{.5\textwidth}
  \centering
  \includegraphics[width=3.25in]{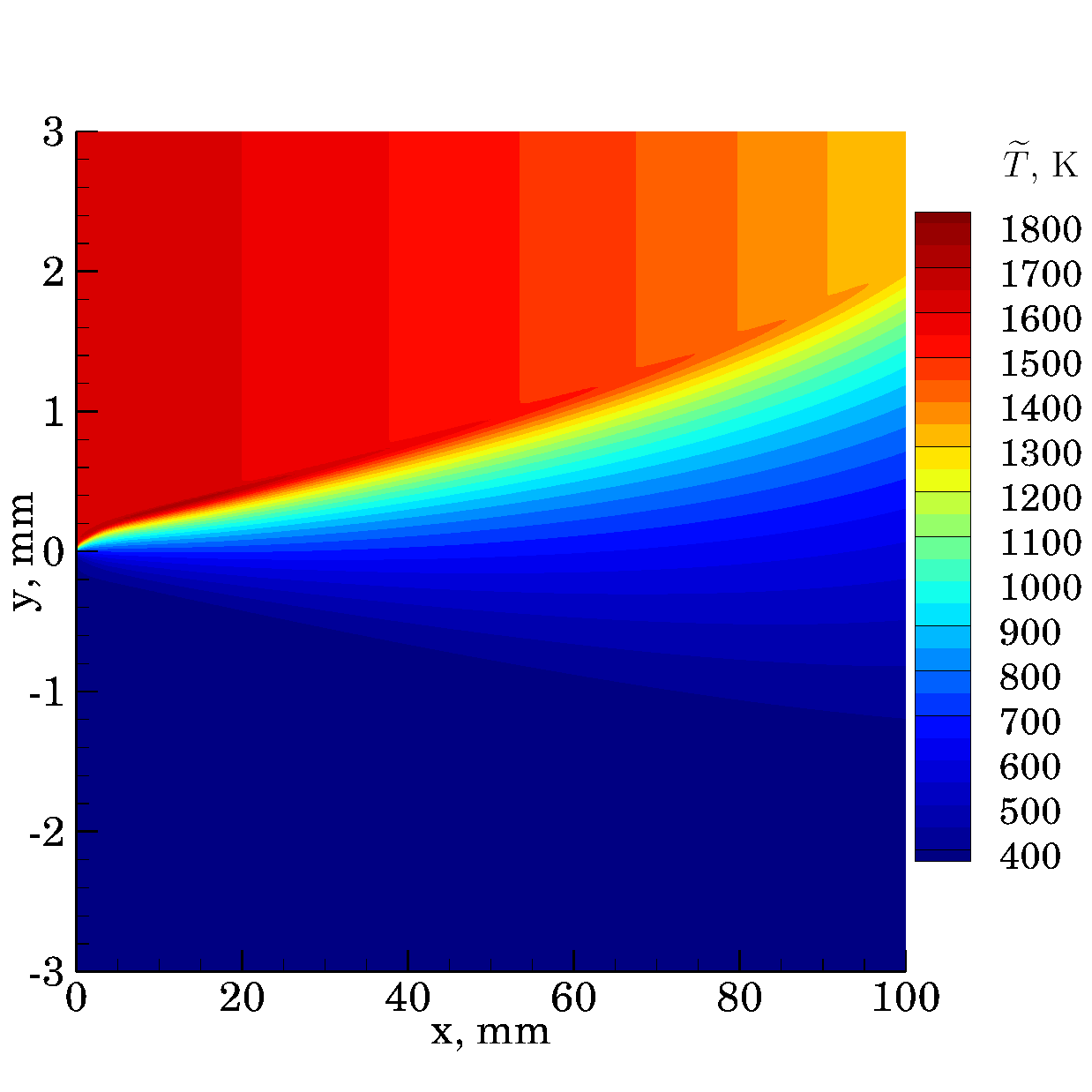}
  \caption{Temperature contour.}
  \label{fig:6r_t_cont}
\end{subfigure}%
\begin{subfigure}{.5\textwidth}
  \centering
  \includegraphics[width=3.25in]{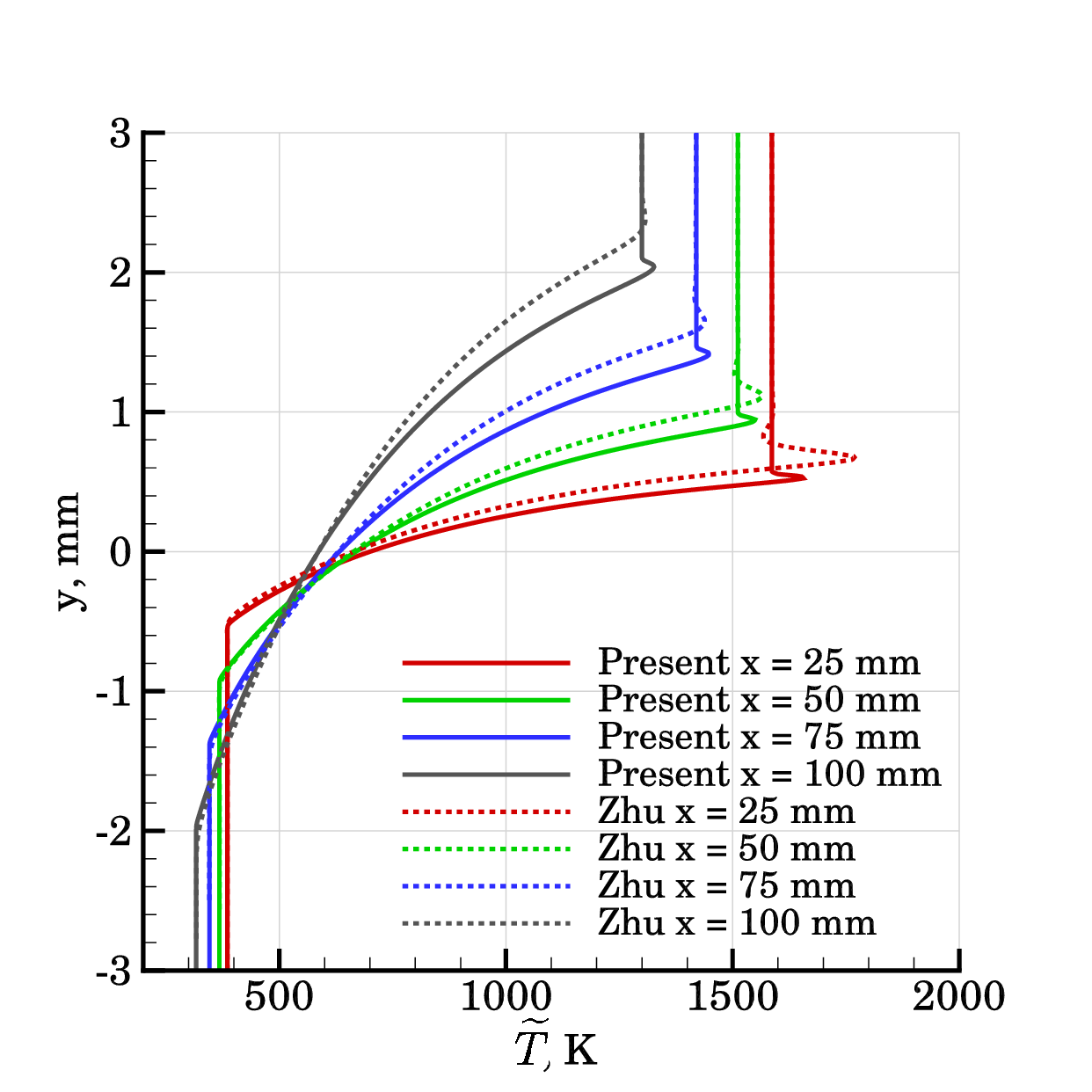}
  \caption{Temperature profiles.}
  \label{fig:6r_profiles}
\end{subfigure}
\caption{Reacting mixing layer results using the vitiated air flamelet libraries.}
\label{fig:6r_t_cont_class}
\end{figure}

Figure \ref{fig:6r_t_cont} shows temperature contours of the vitiated air flame, while Fig. \ref{fig:6r_profiles} plots temperature profiles obtained using the present approach (B-L FPV) and those of Zhu et al. \cite{Zhu2024} (OSK 2D N-S) at four different streamwise locations. The flame behavior in the vitiated air case significantly differs from the pure air case. Notably, the peak flame temperature for the vitiated air case is significantly reduced due to lower $\mathrm{O_2}$ concentration, which weakens the reaction. This results in a higher flame density compared to the pure air case. When comparing the present profiles to those obtained by the OSK approach, a similar trend to what was described in the previous section is observed: the FPV approach reports lower peak temperatures. Distinct mixing layer growths are noted between the B-L and N-S approaches, attributed to the reduced accuracy of the B-L approach in the the region immediately downstream of the edge of the computational domain, near x = 0 mm. Although not shown in a figure, higher peaks of $\mathrm{H_2O}$ and $\mathrm{CO_2}$ are reported in the vitiated air case due to their non-zero air-side freestream concentrations. Additionally, the production of $\mathrm{CO}$ is lower than in the pure air solution, which is indicative of a weaker reaction.

\begin{figure}
\centering
\begin{subfigure}{.5\textwidth}
  \centering
  \includegraphics[width=3.25in]{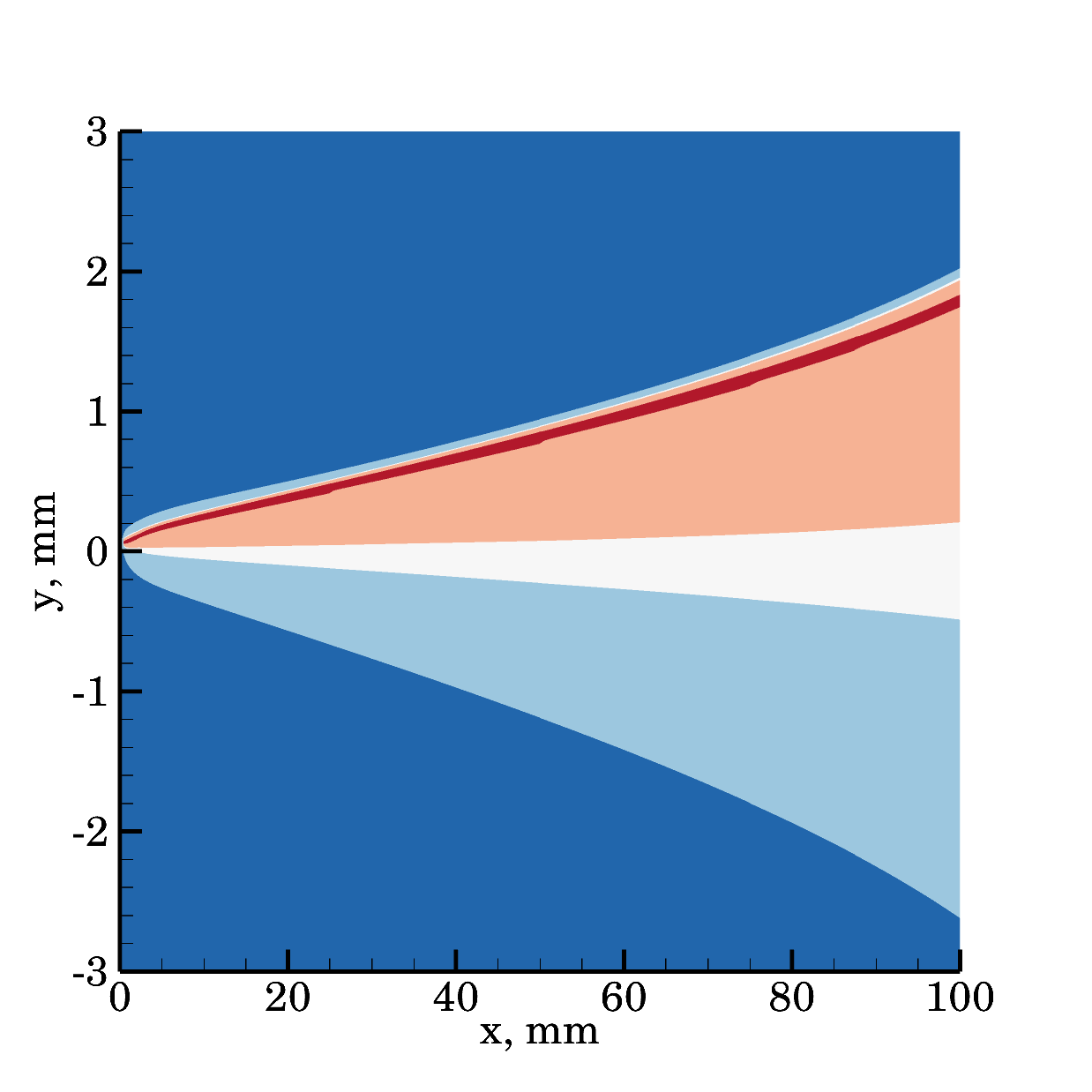}
  \caption{Pure air.}
  \label{fig:pure_air_class}
\end{subfigure}%
\begin{subfigure}{.5\textwidth}
  \centering
  \includegraphics[width=3.25in]{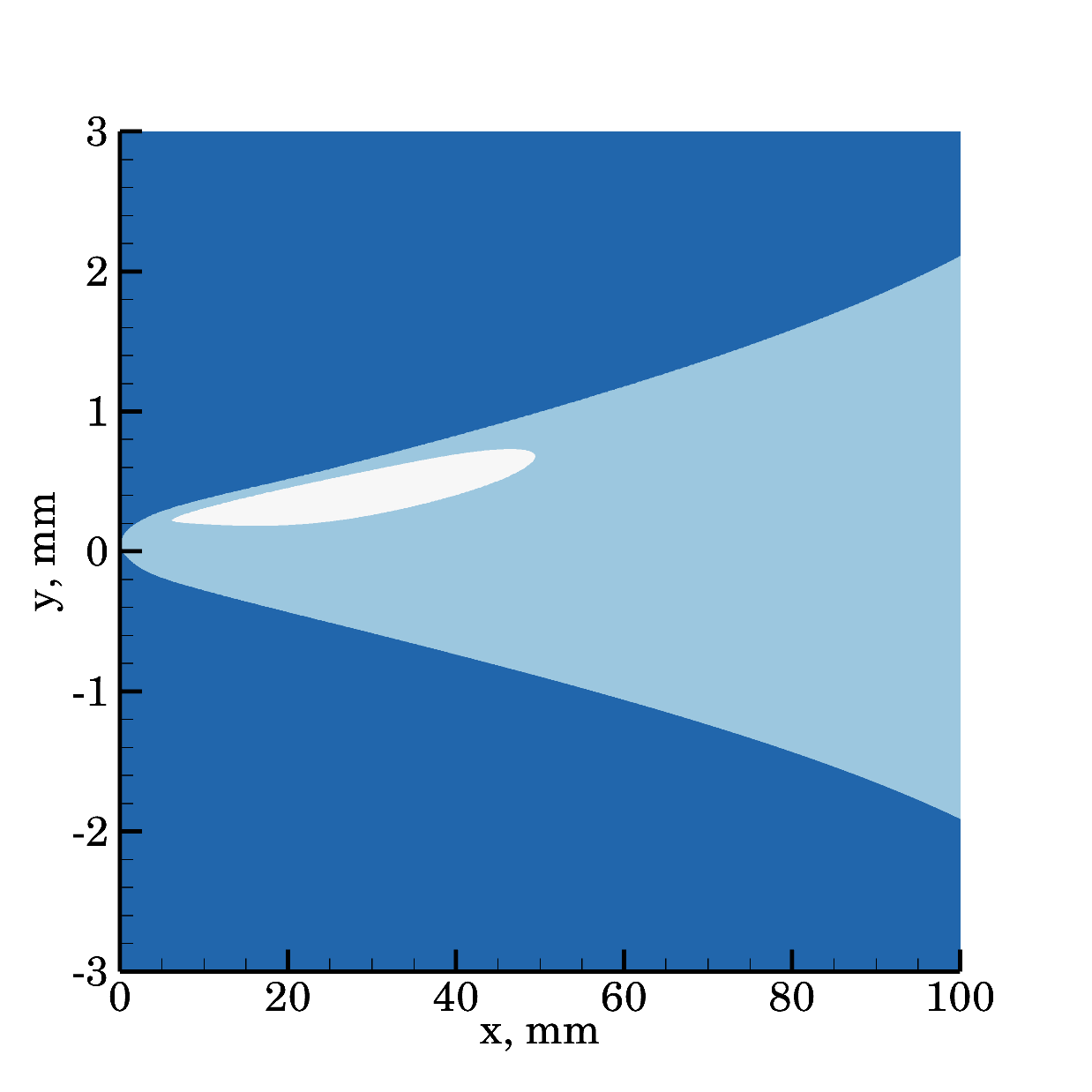}
  \caption{Vitiated air.}
  \label{fig:vit_air_class}
\end{subfigure}
\caption{Flame zone classification. Dark blue: non-burning; light blue: unstable branch; white: around quenching limit; pink: stable branch; and red: closest to equilibrium.}
\label{fig:class}
\end{figure}

Furthermore, apart from reduced temperature due to lower $\mathrm{O_2}$ presence, the development of the vitiated air flame is hindered, with extinction occurring about halfway through the domain length. Further insight into this behavior is provided by Figs. \ref{fig:pure_air_class} and \ref{fig:vit_air_class}, which illustrate the classification of the flame based on the type of sub-grid flamelet encountered at each computational cell according to the "S-shaped" curves. This classification is carried out according to the resolved-scale distribution of the mean progress variable $\widetilde{C}$ (the solution of Eq. \ref{trans_c}). It should be viewed in conjunction with Figs. \ref{fig:s_shaped_pure_air} and \ref{fig:s_shaped_vitiated_air}. In this representation, the dark blue zones outside the mixing layers correspond to non-reacting flamelets, the light blue zones correspond to unstable flamelets, white represents flamelets near and around the quenching limit, pink corresponds to flamelets along the stable branch, and red corresponds to stable flamelets near or at the equilibrium solution (upper left-most points in the "S-shaped" curves).

In the case of pure air, the mixing layer is comprised of flamelets spanning the entirety of the "S-shaped" curve, with stable and equilibrium solutions dominating the flame and reaction zone. This behavior indicates that the flame quickly reaches a "fully burning" state and remains in this state throughout the rest of the computational domain. On the other hand, in the case of vitiated air, stable branch flamelets do not appear. The flame development is hindered, with unstable flamelet solutions dominating both the mixing layer and the reaction zone. A small region of quenching limit solutions appears at around x = 10 mm, indicating impeded ignition compared to the pure air case. Further downstream, this region reverts to unstable solutions, which is a clear indication of extinction.

This hindered development can also be explained by the significant reduction in the quenching limit strain rate ($\chi_{st}$) of the sub-scale flamelet solutions, which are shown to be two orders of magnitude lower than in the pure air composition, as presented in Fig. \ref{fig:s_shaped_new}. Sections \ref{sec:ffcm1_ffcm13_flamelet} and \ref{rs_mech_comp}, the reduced quenching limit strain rate leads to a lower magnitude of the source term of the progress variable, ultimately resulting in delayed reaction development.

Furthermore, these results indicate that vitiated air flames are dominated by unstable solutions. This, coupled with the fact that precise description of the unstable branch is crucial for accurately capturing ignition and extinction phenomena \cite{Pierce2004,Nguyen2018_2}, establishes the grounds for using the FPV approach when treating vitiated air combustion.

\section{Conclusions}
In this study, a compressible extension of the flamelet progress variable (FPV) approach has been proposed for use with LES or RANS analyses of methane and vitiated air combustion under transonic, turbulent, high-pressure, and high-pressure-gradient conditions. The proposed model introduces significant differences from previous compressible FPV approaches, particularly in the determination of resolved-scale temperature, the dependence of sub-grid chemistry on pressure, and the coupling of flamelet libraries to resolved-scale computations. The proposed method has been tested on turbulent, accelerating, combusting, compressible mixing layers using a boundary-layer approximation parabolic solver, coupled with both $k$-$\omega$ and SST turbulence models. The flamelet libraries are generated using the full 38-species FFCM-1 reaction mechanism and a 13-species reduction (FFCM-13). Focus has been placed on the combustion of vitiated air in the turbine-burner concept, seeking greater accuracy in heat release and flameholding, the latter being governed by ignition and extinction phenomena.

Pure air computations using flamelet libraries generated with the FFCM-13 mechanism have been compared against prior boundary-layer solutions and higher fidelity two-dimensional Navier-Stokes solutions, both employing the same resolved-scale OSK combustion model. Results for various flow variables (temperature, density, velocity, and species mass fractions) generally agree both qualitatively and quantitatively with these comparison studies, validating the proposed compressible FPV approach. However, variations due to differences between the FPV and OSK combustion models have been identified. FPV solutions with the more detailed reaction mechanism show faster chemistry compared to the OSK approach, with shorter ignition delays across all cases. Additionally, the FPV method reports significant $\mathrm{CO}$ production in the reaction zone for all cases and a notable reduction in peak flame temperature compared to the OSK solutions. This temperature reduction is attributed to variations in product composition, heat loss to dissociation and radical formation, and turbulence effects on the sub-grid flamelets. Furthermore, FPV results show to have stronger sensitivity to background pressure, justifying the use of the pressure dimension for the flamelet libraries. Both FPV and OSK solutions captured significant amounts of $\mathrm{O_2}$ entrainment from the air-side to the fuel-side due to flame ignition delay, but a significant shortcoming of the FPV approach has been identified in terms of the lack of diffusion of the entrained $\mathrm{O_2}$, affecting the mixing layer structure.

Correlations are identified between portions of the flamelet library results and domains of the flow field based on composition (progress variable value). These correlations provide insights into the structure and development of the flames. An important finding is that vitiated air flames are dominated by unstable flamelet solutions, resulting in a weak flame with substantially lower peak temperature and impeded development, struggling to persist without quenching.

Resolved-scale computations comparing flamelet libraries generated with both reduced and full mechanisms show that differences in sub-grid chemistry, particularly flammability limits (quenching strain rates), temperature, and species production, carry over to the resolved scale. These differences lead to variations in predictions for ignition, temperature, and species production at the resolved-scale. These comparisons reveal that in the FPV approach, chemistry, rather than resolved-scale mechanics, governs the flamelet strain rate.

While experimental validation is currently unavailable, assuming the reliability of FFCM-1 mechanism, it is argued that the FPV model using the full reaction mechanism, with its more comprehensive representation of chemical kinetics, should provide greater accuracy in capturing the combustion process compared to the FFCM-13 model. By the same reasoning, any mechanism used within the FPV approach is expected to outperform OSK models, which, due to current computational limitations, cannot match the level of chemical kinetic detail of the FPV models. Given the significant differences in temperature predictions between the FPV models and the OSK approach, and considering the critical importance of accurate temperature predictions in turbine burners, where heat transfer to the turbine blades is a key factor, the use of an FPV model is strongly justified for simulations in this application.

\section*{Acknowledgments}
The research was supported by the Office of Naval Research through Grant N00014-22-1-2467 with Dr. Steven Martens as program manager. Professor Heinz Pitsch of RWTH Aachen University is acknowledged for providing us access to the FlameMaster code.

\bibliography{library}

\end{document}